\documentclass[aps,prd,onecolumn,nofootinbib,superscriptaddress]{revtex4-2}
\usepackage[utf8]{inputenc}
\usepackage{amsmath,amssymb}
\usepackage{physics}
\usepackage{graphicx}
\usepackage{xcolor}
\usepackage{appendix}
\usepackage{placeins}

\begin{document}

\title{Dynamical systems analysis of unimodular cosmology in $D=4+d$ dimensions}

\author{Vel\'asquez-Toribio, A. M.}
\email{alan.toribio@ufes.br}
\affiliation{Center for Astrophysics and Cosmology of the Esp\'irito Santo Federal University (UFES), 29075-910 Vit\'oria - ES, Brazil}

\author{Fabris, J. C.}
\email{julio.fabris@cosmo-ufes.org}
\affiliation{Center for Astrophysics and Cosmology of the Esp\'irito Santo Federal University (UFES), 29075-910 Vit\'oria - ES, Brazil}

\begin{abstract}
We investigate the effective four-dimensional cosmology induced by
unimodular gravity in $D=4+d$ dimensions, where the internal
extra-dimensional volume is encoded in a scalar degree of freedom. After
dimensional reduction, we show that the resulting FLRW equations admit a
natural autonomous formulation whose phase-space structure differs
qualitatively from that of general relativity. In the vacuum sector, the
reduced system exhibits a continuous family of finite equilibrium points,
$\lambda=dH$, together with well-defined asymptotic Poincar\'e directions.
In the matter sector, we focus on the five-dimensional case $d=1$ and use
the reduced Bianchi relation as the consistency condition that links the
ordinary matter component to the internal-volume degree of freedom. The
system is then closed by adopting the minimal higher-dimensional
conservation prescription, according to which matter is diluted by both
the external volume and the internal-volume modulus. This leads to a
reduced matter--geometry dynamics with isolated critical points and a
globally organized compactified flow. Numerical examples illustrate how
the internal-volume degree of freedom affects the background evolution and
the global phase-space structure. The comparison with $\Lambda$CDM is used
only as a benchmark, while a full observational analysis and more general
matter--geometry exchange prescriptions are left for future work.
\end{abstract}

\maketitle

\section{Introduction}

One of the central problems of contemporary cosmology is the observed accelerated expansion of the Universe. Evidence from type Ia supernovae, cosmic microwave background anisotropies, and large-scale structure indicates that the late-time cosmic dynamics is currently dominated by a component with negative effective pressure. Within the standard cosmological model, this phenomenon is usually described by a cosmological constant, leading to the successful $\Lambda$CDM scenario \cite{Peebles2003,Clifton2012,Perivolaropoulos2022,Martin2012}. However, despite its excellent phenomenological performance, this framework is still accompanied by deep conceptual difficulties, most notably the cosmological constant problem and the coincidence problem \cite{Zimdahl2014}. These difficulties motivate the search for alternative formulations of gravitation in which the effective cosmological dynamics may emerge from geometry itself, or from a different implementation of the gravitational degrees of freedom, instead of being attributed solely to an \textit{ad hoc} dark-energy sector.

In this context, unimodular gravity has long been regarded as an interesting alternative viewpoint. In its simplest form, the theory is based on fixing the determinant of the metric, or equivalently the spacetime volume element, which modifies the variational structure of general relativity and leads naturally to the traceless part of Einstein's equations. Historically, the roots of this idea go back to Einstein's early considerations on restricted covariance \cite{Einstein1919}, while its modern formulation was developed much later in a more systematic way. In particular, the work of Anderson and Finkelstein provided one of the first explicit modern formulations of unimodular relativity \cite{AndersonFinkelstein1971,vanderBijNgDam1982,BuchmullerDragon1988}, and the later approaches of Henneaux and Teitelboim and of Unruh played an important role in establishing its canonical and covariant interpretation \cite{HenneauxTeitelboim1989, Unruh1989,NgVanDam1991, FinkelsteinGaliautdinovBaugh2001, Alvarez2005}. In this framework, the cosmological constant appears not as a parameter inserted directly in the action, but rather as an integration constant of the field equations. For this reason, unimodular gravity has often been revisited in discussions of the vacuum-energy problem, quantum gravity, and cosmology \cite{AlvarezFaedo2007, Smolin2009, FiolGarriga2010, LopezVillarejo2011, EllisEtAl2011, JainEtAl2012, GaoEtAl2014, PadillaSaltas2015, AlvarezEtAl2015, BufaloOksanenTureanu2015, JossetPerezSudarsky2017, BonderCorral2018,GarciaAspeitiaEtAl2019, CorralCruzGonzalez2020,Salvio2024}.

Over the years, the unimodular proposal has been explored from several complementary directions. At the classical level, its relation to general relativity and the extent to which both theories are dynamically equivalent have been carefully discussed. At the same time, unimodular models have also been studied in inflationary scenarios, cosmological perturbation theory, and late-time cosmology, as well as in extensions in which the standard conservation law is relaxed or reformulated \cite{GarciaAspeitiaEtAl2021,Kaczmarek2024,AlvarezVelascoAja2023,CarballoRubio2022,FabrisFallerKerner2026,NojiriOdintsovOikonomou2016,LinaresCedenoEtAl2023,AlcanizEtAl2012,OdintsovOikonomou2016,PlazaLeon2025}. More recently, there has been renewed interest in clarifying conceptual issues of the theory and in confronting unimodular cosmological models with observational data \cite{PiccirilliLeon2023}. Altogether, this body of work shows that, even when the theory reproduces general relativity in many circumstances, its formulation suggests a distinct conceptual perspective and may lead to nontrivial cosmological realizations once additional structures are incorporated.

A particularly natural extension in this direction is the inclusion of extra dimensions. Higher-dimensional models remain an important arena in gravitational physics since they arise in Kaluza-Klein constructions and in broader attempts to unify gravity with other interactions. After dimensional reduction, the internal sector typically gives rise to effective scalar degrees of freedom in four dimensions, associated with the geometry or volume of the compact space. These degrees of freedom can modify the expansion history, affect the effective matter content, and alter the phase-space structure of cosmological solutions. For this reason, combining unimodular gravity with extra dimensions opens a promising route to obtain nontrivial effective cosmological dynamics directly from geometry.

This possibility has already begun to receive attention in the literature. In particular, recent studies have investigated how the unimodular condition can be implemented in five-dimensional Kaluza-Klein settings and how the resulting dimensional reduction leads to effective four-dimensional cosmological equations involving scalar contributions from the extra-dimensional sector \cite{Erdem2026,Jirousek2023,FabrisKerner2024,FabrisKerner2025}. These developments suggest that the combination of unimodular gravity and higher dimensions is not only mathematically consistent, but also cosmologically rich enough to justify a systematic analysis of its dynamical content.

The main purpose of this work is not simply to apply dynamical-systems methods to a higher-dimensional cosmological model, but to identify the specific dynamical imprint produced by the combination of unimodular gravity and extra dimensions after dimensional reduction. In particular, we show that the reduced theory possesses a nontrivial equilibrium structure already in vacuum, including a continuous critical set, and that in the five-dimensional matter-coupled case the effective Bianchi relation leads naturally to a coupled matter-geometry system with its own reduced phase-space organization. This provides a global characterization of the cosmological dynamics associated with the unimodular extra-dimensional sector, beyond a mere comparison with the standard of the General Relativity case.

This paper is organized as follows. In Sec.~II we derive the effective four-dimensional field equations starting from the unimodular theory in $D=4+d$ dimensions. In Sec.~III we specialize the model to a flat FLRW background and obtain the corresponding autonomous systems in the vacuum and matter sectors. In Sec.~IV we analyze the equilibrium structure and the asymptotic Poincar\'e directions of the model. In Sec.~V we present numerical illustrations of the cosmological trajectories and phase portraits and in Sec. VI we summarize our results and comment on their physical interpretation.

\section{The field equations for unimodular theory}

We begin with the unimodular field equations in $D=4+d$ dimensions,
\begin{eqnarray}
\tilde R_{AB}-\frac{1}{D}\tilde g_{AB}\tilde R
=
8\pi G\left(\tilde T_{AB}-\frac{1}{D}\tilde g_{AB}\tilde T\right),
\label{eq:uniD}
\end{eqnarray}
together with the corresponding divergence condition.

In order to obtain an effective four-dimensional description, we assume
that the higher-dimensional manifold splits into a four-dimensional
spacetime and a $d$-dimensional internal Euclidean sector. We then adopt
the metric ansatz
\begin{eqnarray}
ds_D^2
=
dt^2-a^2(t)\delta_{ij}dx^i dx^j-b^2(t)\delta_{ab}dx^a dx^b,
\label{eq:metricD}
\end{eqnarray}
where $i,j=1,2,3$ and $a,b=1,\ldots,d$. The function $b(t)$ is the scale
factor of the internal space.

Since the physical effect of the extra dimensions is associated with the
evolution of the internal volume, it is convenient to introduce the scalar
quantity
\begin{eqnarray}
\psi \equiv b^d.
\label{eq:psi_def}
\end{eqnarray}
Thus, $\psi$ represents the volume modulus of the compact internal sector.

For the metric (\ref{eq:metricD}), the decomposition $D=4+d$ yields
\begin{eqnarray}
\tilde R_{\mu\nu}
=
R_{\mu\nu}
-
d\,\frac{b_{;\mu;\nu}}{b},
\label{eq:Rmunu_split}
\end{eqnarray}
\begin{eqnarray}
\tilde R_{ab}
=
\left(
b\Box b+(d-1)b_{;\rho}b^{;\rho}
\right)\delta_{ab},
\label{eq:Rab_split}
\end{eqnarray}
and
\begin{eqnarray}
\tilde R
=
R
-
2d\frac{\Box b}{b}
-
d(d-1)\frac{b_{;\rho}b^{;\rho}}{b^2}.
\label{eq:Rscalar_split}
\end{eqnarray}

We assume that matter is confined to the four-dimensional sector, so that
\begin{eqnarray}
\tilde T_{\mu\nu}=T_{\mu\nu},
\qquad
\tilde T_{ab}=0,
\qquad
\tilde T=T.
\label{eq:matter_split}
\end{eqnarray}
Substituting Eqs.~(\ref{eq:Rmunu_split})--(\ref{eq:Rscalar_split}) and
(\ref{eq:matter_split}) into Eq.~(\ref{eq:uniD}), the $\mu\nu$ components
become
\begin{eqnarray}
R_{\mu\nu}
-
\frac{1}{d+4}g_{\mu\nu}R
=
8\pi G\left(T_{\mu\nu}-\frac{1}{d+4}g_{\mu\nu}T\right)
-
\frac{d(d-1)}{d+4}g_{\mu\nu}\frac{b_{;\rho}b^{;\rho}}{b^2}
+
\frac{d}{b}\left(
b_{;\mu;\nu}-\frac{2}{d+4}g_{\mu\nu}\Box b
\right).
\label{eq:reduced_b}
\end{eqnarray}

The extra-dimensional $(ab)$ components provide an additional scalar
relation. Since $\tilde T_{ab}=0$, Eq.~(\ref{eq:uniD}) gives
\begin{eqnarray}
\tilde R_{ab}
-
\frac{1}{d+4}\tilde g_{ab}\tilde R
=
-\frac{8\pi G}{d+4}\tilde g_{ab}T .
\label{eq:abstart}
\end{eqnarray}
Using Eqs.~(\ref{eq:Rab_split}) and (\ref{eq:Rscalar_split}) in
Eq.~(\ref{eq:abstart}), and taking into account that
$\tilde g_{ab}=-b^2\delta_{ab}$, we obtain, after dividing by the common
factor $\delta_{ab}$,
\begin{eqnarray}
b\Box b+(d-1)b_{;\rho}b^{;\rho}
+\frac{b^2}{d+4}
\left[
R
-
2d\frac{\Box b}{b}
-
d(d-1)\frac{b_{;\rho}b^{;\rho}}{b^2}
\right]
=
\frac{8\pi G}{d+4}b^2 T.
\label{eq:ab_intermediate}
\end{eqnarray}
Dividing by $b^2$ and collecting terms, one finds
\begin{eqnarray}
(d-4)\frac{\Box b}{b}
-
4(d-1)\frac{b_{;\rho}b^{;\rho}}{b^2}
=
R-8\pi G T,
\label{eq:ab_rearranged}
\end{eqnarray}
or, equivalently,
\begin{eqnarray}
\frac{\Box b}{b}
-
\frac{4(d-1)}{d-4}\frac{b_{;\rho}b^{;\rho}}{b^2}
=
\frac{1}{d-4}\left(R-8\pi G T\right),
\label{eq:b_scalar_eq}
\end{eqnarray}
which is valid for $d\neq 4$.

It is now convenient to rewrite the reduced theory in terms of the
internal-volume scalar $\psi=b^d$. Using
\begin{eqnarray}
\frac{\psi_{;\mu}}{\psi}
=
d\,\frac{b_{;\mu}}{b},
\qquad
\frac{\psi_{;\mu;\nu}}{\psi}
=
d\,\frac{b_{;\mu;\nu}}{b}
+
d(d-1)\frac{b_{;\mu}b_{;\nu}}{b^2},
\label{eq:psi_relations}
\end{eqnarray}
the effective four-dimensional equations can be recast as
\begin{eqnarray}
R_{\mu\nu}
=
8\pi G T_{\mu\nu}
+
\frac{1-d}{d}\frac{\psi_{;\mu}\psi_{;\nu}}{\psi^2}
+
\frac{1}{\psi}
\left(
\psi_{;\mu;\nu}
-
\frac{1}{d}g_{\mu\nu}\Box\psi
\right).
\label{eq:eff_tensor}
\end{eqnarray}

From Eq.~(\ref{eq:b_scalar_eq}), one also obtains
\begin{eqnarray}
\frac{\Box\psi}{\psi}
=
\frac{1-d}{4-d}\frac{\psi_{;\rho}\psi^{;\rho}}{\psi^2}
+
\frac{d}{4-d}\left(8\pi G T-R\right),
\label{eq:eff_scalar}
\end{eqnarray}
which is the form most convenient for the cosmological analysis.

At this stage, it is important to examine the consistency of the reduced
field equations under covariant differentiation. Since
\begin{eqnarray}
\nabla^\mu G_{\mu\nu}=0,
\qquad
G_{\mu\nu}=R_{\mu\nu}-\frac{1}{2}g_{\mu\nu}R,
\label{eq:bianchi_einstein}
\end{eqnarray}
one has the contracted Bianchi identity
\begin{eqnarray}
\nabla^\mu R_{\mu\nu}
=
\frac{1}{2}R_{;\nu}.
\label{eq:contracted_bianchi}
\end{eqnarray}
Applying $\nabla^\mu$ to Eq.~(\ref{eq:eff_tensor}), and using
Eq.~(\ref{eq:eff_scalar}), one obtains a generalized consistency relation
for the effective four-dimensional source. In the intermediate steps, the
commutation of covariant derivatives acting on the scalar field is required,
namely,
\begin{eqnarray}
\psi_{;\mu;\nu}{}^{;\mu}
=
(\Box\psi)_{;\nu}
+
R_{\mu\nu}\psi^{;\mu}.
\label{eq:commutator_scalar}
\end{eqnarray}
After a straightforward but lengthy calculation, the result can be written
as
\begin{eqnarray}
\frac{d + 2}{2(4 - d)}R_{;\nu} = 8\pi G\biggr\{{T^\mu_\nu}_{;\mu} + \frac{\psi^{;\mu}}{\psi}T_{\mu\nu} - \frac{1 - d}{4 - d}T_{;\nu}\biggl\} \nonumber\\
+ \frac{(1 - d)}{d}\frac{d + 2}{4 - d}\frac{\psi^{;\mu}}{\psi}\biggr(\frac{\psi_{;\mu}}{\psi}\biggl)_{;\nu}.
\label{eq:bianchi_general}
\end{eqnarray}
In order to obtain this expression, the following relation has been used:
\begin{eqnarray}
\psi^{;\mu}_{;\nu;\mu} = (\Box\psi)_{;\nu} + R_{\mu\nu}\psi^{;\mu}.
\end{eqnarray}

Equation~(\ref{eq:bianchi_general}) is the general form of the effective
Bianchi relation in the reduced theory. It shows that, once the
extra-dimensional sector is encoded in the scalar field $\psi$, the
divergence properties of the four-dimensional matter tensor are not
isolated from the geometric scalar contribution. In other words, the
consistency condition implied by the Bianchi identities constrains the total
effective source appearing on the right-hand side of Eq.~(\ref{eq:eff_tensor}),
rather than the ordinary matter sector alone.

A particularly simple situation arises for $d=1$. In this case, all terms
proportional to $(1-d)$ vanish identically, and Eq.~(\ref{eq:bianchi_general})
reduces to
\begin{eqnarray}
\frac{1}{2}R_{;\nu}
=
8\pi G
\left(
T^{\mu}{}_{\nu;\mu}
+
\frac{\psi^{;\mu}}{\psi}T_{\mu\nu}
\right).
\label{eq:bianchi_d1}
\end{eqnarray}
Thus, in the five-dimensional case the effective Bianchi relation takes a
particularly compact form, with the scalar sector entering only through the
mixed coupling term $(\psi_{;\mu}/\psi)T^{\mu}{}_{\nu}$.

Equations~(\ref{eq:eff_tensor}), (\ref{eq:eff_scalar}), and
(\ref{eq:bianchi_general}) provide the complete covariant formulation of
the reduced theory in terms of the single scalar field $\psi$. They
constitute the appropriate starting point for the cosmological reduction to
be performed in the next section, where the formalism will be specialized to
a spatially flat FLRW spacetime and the corresponding autonomous systems
will be constructed.

\section{Cosmology}

We now specialize the effective equations (\ref{eq:eff_tensor}) and
(\ref{eq:eff_scalar}) to a spatially flat FLRW spacetime,
\begin{eqnarray}
ds^2
=
dt^2-a^2(t)\left(dx^2+dy^2+dz^2\right),
\qquad
H\equiv \frac{\dot a}{a},
\label{eq:flrw_metric}
\end{eqnarray}
and assume that the scalar associated with the internal volume is
homogeneous,
\begin{eqnarray}
\psi=\psi(t).
\label{eq:psi_hom}
\end{eqnarray}
Since $\psi$ has already been defined as $\psi=b^d$, no new scalar degree
of freedom is introduced at this stage.

We use the metric signature $+---$ and the curvature convention for which
the nonvanishing FLRW curvature components are
\begin{eqnarray}
R_{00}
=
-3(\dot H+H^2),
\qquad
R_{ij}
=
-(\dot H+3H^2)g_{ij},
\qquad
R
=
-6(\dot H+2H^2).
\label{eq:flrw_curvature}
\end{eqnarray}
The sign of the Ricci scalar in Eq.~(\ref{eq:flrw_curvature}) follows
from the contraction of $R_{\mu\nu}$ with the same metric convention.
This sign will be important below, when the reduced Bianchi relation is
used.

For the homogeneous scalar $\psi(t)$ one has
\begin{eqnarray}
\psi_{;\rho}\psi^{;\rho}
=
\dot\psi^2,
\qquad
\frac{\Box\psi}{\psi}
=
\frac{\ddot\psi}{\psi}
+
3H\frac{\dot\psi}{\psi},
\qquad
\psi_{;i;j}
=
H g_{ij}\dot\psi .
\label{eq:scalar_flrw_relations}
\end{eqnarray}
It is convenient to introduce
\begin{eqnarray}
\lambda
\equiv
\frac{\dot\psi}{\psi},
\qquad
\frac{\ddot\psi}{\psi}
=
\dot\lambda+\lambda^2 .
\label{eq:lambda_def}
\end{eqnarray}

\subsection{Vacuum sector}

In the vacuum sector, $T_{\mu\nu}=0$, the $00$ and $ij$ components of
Eq.~(\ref{eq:eff_tensor}) reduce to
\begin{eqnarray}
3(\dot H+H^2)
=
-
\frac{1-d}{d}\lambda^2
-
\frac{d-1}{d}\frac{\ddot\psi}{\psi}
+
\frac{3}{d}H\lambda ,
\label{eq:vac_00}
\end{eqnarray}
and
\begin{eqnarray}
\dot H+3H^2
=
\frac{1}{d}\frac{\ddot\psi}{\psi}
+
\frac{3-d}{d}H\lambda .
\label{eq:vac_ij}
\end{eqnarray}
Using Eq.~(\ref{eq:lambda_def}) in
Eqs.~(\ref{eq:vac_00}) and (\ref{eq:vac_ij}), and eliminating
$\dot\lambda$, one obtains the autonomous vacuum system
\begin{eqnarray}
\dot H
=
-
\frac{1}{d(d+2)}
\left[
3d^2H^2
+
(1-d)\lambda^2
+
d(d-4)H\lambda
\right],
\label{eq:Hdot_vac}
\end{eqnarray}
\begin{eqnarray}
\dot\lambda
=
\frac{1}{d+2}
\left[
6dH^2
-
3\lambda^2
+
3(d-2)H\lambda
\right].
\label{eq:lambdadot_vac}
\end{eqnarray}
This quadratic system governs the vacuum cosmological dynamics of the
reduced unimodular model.

\subsection{Matter coupled to the geometric sector}

We now include a four-dimensional perfect fluid,
\begin{eqnarray}
T_{\mu\nu}
=
(\rho+p)u_\mu u_\nu-pg_{\mu\nu},
\label{eq:perfect_fluid}
\end{eqnarray}
with $u^\mu u_\mu=1$. For the FLRW background, Eq.~(\ref{eq:eff_tensor})
becomes
\begin{eqnarray}
3(\dot H+H^2)
=
-8\pi G\rho
-
\frac{1-d}{d}\lambda^2
-
\frac{d-1}{d}\frac{\ddot\psi}{\psi}
+
\frac{3}{d}H\lambda ,
\label{eq:matter_00}
\end{eqnarray}
and
\begin{eqnarray}
\dot H+3H^2
=
8\pi G p
+
\frac{1}{d}\frac{\ddot\psi}{\psi}
+
\frac{3-d}{d}H\lambda .
\label{eq:matter_ij}
\end{eqnarray}
Using Eq.~(\ref{eq:lambda_def}) and eliminating $\dot\lambda$, one finds
\begin{eqnarray}
\dot H
=
-
\frac{1}{d(d+2)}
\left[
3d^2H^2
+
8\pi Gd\rho
-
8\pi Gd(d-1)p
+
(1-d)\lambda^2
+
d(d-4)H\lambda
\right],
\label{eq:Hdot_matter_general_p}
\end{eqnarray}
\begin{eqnarray}
\dot\lambda
=
\frac{1}{d+2}
\left[
6dH^2
-
8\pi Gd\rho
-
24\pi Gd p
-
3\lambda^2
+
3(d-2)H\lambda
\right].
\label{eq:lambdadot_matter_general_p}
\end{eqnarray}
In what follows we specialize to pressureless matter, $p=0$. In this case
the previous equations reduce to
\begin{eqnarray}
\dot H
=
-
\frac{1}{d(d+2)}
\left[
3d^2H^2
+
8\pi Gd\rho
+
(1-d)\lambda^2
+
d(d-4)H\lambda
\right],
\label{eq:Hdot_matter_general}
\end{eqnarray}
\begin{eqnarray}
\dot\lambda
=
\frac{1}{d+2}
\left[
6dH^2
-
8\pi Gd\rho
-
3\lambda^2
+
3(d-2)H\lambda
\right].
\label{eq:lambdadot_matter_general}
\end{eqnarray}

At this stage, the matter evolution is not fixed by
Eqs.~(\ref{eq:Hdot_matter_general}) and
(\ref{eq:lambdadot_matter_general}) alone. The reduced Bianchi relation
derived in the previous section must be interpreted as an integrability
condition for the total effective source, rather than as an independent
conservation law for the ordinary four-dimensional matter sector.

In the special case $d=1$, Eq.~(\ref{eq:bianchi_d1}) gives
\begin{eqnarray}
\frac{1}{2}R_{;\nu}
=
8\pi G
\left(
T^{\mu}{}_{\nu;\mu}
+
\frac{\psi_{;\mu}}{\psi}T^{\mu}{}_{\nu}
\right).
\label{eq:bianchi_d1_cosmo}
\end{eqnarray}
For a homogeneous FLRW background, only the $\nu=0$ component is
nontrivial. Using
\begin{eqnarray}
T^{\mu}{}_{0;\mu}
=
\dot\rho+3H(\rho+p),
\qquad
\frac{\psi_{;\mu}}{\psi}T^{\mu}{}_{0}
=
\frac{\dot\psi}{\psi}\rho
=
\lambda\rho ,
\label{eq:matter_divergence_flrw}
\end{eqnarray}
one obtains
\begin{eqnarray}
\frac{1}{2}\dot R
=
8\pi G
\left[
\dot\rho+3H(\rho+p)+\lambda\rho
\right].
\label{eq:bianchi_flrw_general}
\end{eqnarray}
For pressureless matter, $p=0$, this becomes
\begin{eqnarray}
\dot\rho+3H\rho+\lambda\rho
=
\frac{\dot R}{16\pi G}.
\label{eq:bianchi_flrw_dust}
\end{eqnarray}
This equation shows that the nonstandard evolution of the
four-dimensional matter density is tied to the time variation of the Ricci
scalar. It does not, by itself, provide an independent evolution equation
for $\rho$.

\subsection{Matter closure and exchange function}

The previous discussion shows that an additional physical prescription is
needed in order to close the matter-coupled system. A general way of
parametrizing the possible matter--geometry exchange is to write
\begin{eqnarray}
\dot\rho+3H(\rho+p)+\lambda\rho
=
Q ,
\label{eq:general_Q_closure}
\end{eqnarray}
where $Q$ is an effective exchange function. Combining this expression
with Eq.~(\ref{eq:bianchi_flrw_general}) gives
\begin{eqnarray}
\dot R
=
16\pi G Q.
\label{eq:Rdot_Q}
\end{eqnarray}
Thus, the exchange function controls the evolution of the Ricci scalar.
Different choices of $Q$ correspond to different physical assumptions
about the interaction between the ordinary matter sector, the internal
volume modulus, and the curvature contribution.

Three cases are particularly useful. First, the minimal
higher-dimensional conservation closure is
\begin{eqnarray}
Q=0.
\label{eq:Q_zero}
\end{eqnarray}
In this case,
\begin{eqnarray}
\dot\rho+3H(\rho+p)+\lambda\rho=0,
\label{eq:HD_conservation_closure}
\end{eqnarray}
and Eq.~(\ref{eq:Rdot_Q}) implies
\begin{eqnarray}
\dot R=0.
\label{eq:R_constant}
\end{eqnarray}
For pressureless matter this closure gives
\begin{eqnarray}
\dot\rho+3H\rho+\lambda\rho=0,
\qquad
\rho a^3\psi=\mathrm{constant}.
\label{eq:rho_scaling_minimal}
\end{eqnarray}
Therefore, the matter density is diluted by both the ordinary
three-dimensional volume $a^3$ and the internal volume modulus $\psi$.
This closure is naturally motivated by the covariant conservation of the
higher-dimensional matter source with vanishing internal pressure.

Second, one may impose the usual four-dimensional conservation law,
\begin{eqnarray}
\dot\rho+3H(\rho+p)=0.
\label{eq:standard_4D_conservation}
\end{eqnarray}
In the notation of Eq.~(\ref{eq:general_Q_closure}), this corresponds to
\begin{eqnarray}
Q=\lambda\rho .
\label{eq:Q_standard_4D}
\end{eqnarray}
In this case the matter density follows the standard four-dimensional
dilution law, while the reduced Bianchi relation implies
\begin{eqnarray}
\dot R=16\pi G\lambda\rho .
\label{eq:Rdot_standard_4D}
\end{eqnarray}

Third, one may keep $Q\neq 0$ as a phenomenological interaction term. This
more general possibility describes a nontrivial exchange between the
matter sector, the internal-volume scalar, and the curvature scalar. Such
models may be useful for future observational studies, but they require a
separate physical or phenomenological prescription for $Q$.
In the present work we adopt the simplest and most conservative choice,
namely the minimal closure
\begin{eqnarray}
Q=0.
\label{eq:adopted_Q_zero}
\end{eqnarray}

The choice is restrictive in the sense that it fixes the
four-dimensional Ricci scalar as an integration constant, $\dot R=0$.
Thus it selects the minimal sector of the reduced theory, rather than the
most general matter--geometry exchange allowed by the Bianchi relation.
For $R=0$ the background kinematics is radiation-like, whereas a nonzero
constant $R$ plays the role of an effective curvature scale analogous to
a cosmological-constant contribution. In the present work we use this
minimal closure only as a reference case; more general choices with
$Q\neq0$, corresponding to a dynamical Ricci scalar, are left for future
study.

\subsection{The minimally closed coupled system for $d=1$}

The case $d=1$ is particularly simple. From
Eqs.~(\ref{eq:Hdot_matter_general}) and
(\ref{eq:lambdadot_matter_general}), one obtains
\begin{eqnarray}
\dot H
=
-H^2
-
\frac{8\pi G}{3}\rho
+
H\lambda ,
\label{eq:Hdot_d1}
\end{eqnarray}
\begin{eqnarray}
\dot\lambda
=
2H^2
-
\frac{8\pi G}{3}\rho
-
\lambda^2
-
H\lambda .
\label{eq:lambdadot_d1}
\end{eqnarray}
With the curvature convention adopted in Eq.~(\ref{eq:flrw_curvature}),
the Ricci scalar is
\begin{eqnarray}
R
=
-6(\dot H+2H^2).
\label{eq:R_flrw_correct}
\end{eqnarray}
Using Eq.~(\ref{eq:Hdot_d1}), this becomes
\begin{eqnarray}
R
=
-6H^2
-
6H\lambda
+
16\pi G\rho .
\label{eq:R_d1_correct}
\end{eqnarray}
Differentiating Eq.~(\ref{eq:R_d1_correct}) and using
Eqs.~(\ref{eq:Hdot_d1}) and (\ref{eq:lambdadot_d1}), one obtains
\begin{eqnarray}
\frac{\dot R}{16\pi G}
=
\dot\rho+3H\rho+\lambda\rho .
\label{eq:Rdot_identity}
\end{eqnarray}
Therefore, Eq.~(\ref{eq:bianchi_flrw_dust}) becomes an identity once the
field equations are used. It does not provide an independent evolution
equation for $\rho$. The evolution of the matter density is fixed only
after the matter closure has been specified.

With the minimal choice $Q=0$, the pressureless matter equation is
\begin{eqnarray}
\dot\rho
=
-3H\rho-\lambda\rho .
\label{eq:rho_dot_minimal}
\end{eqnarray}
Thus, the closed matter--geometry system for $d=1$ is
\begin{eqnarray}
\dot H
=
-H^2
-
\frac{8\pi G}{3}\rho
+
H\lambda ,
\label{eq:closed_H}
\end{eqnarray}
\begin{eqnarray}
\dot\lambda
=
2H^2
-
\frac{8\pi G}{3}\rho
-
\lambda^2
-
H\lambda ,
\label{eq:closed_lambda}
\end{eqnarray}
\begin{eqnarray}
\dot\rho
=
-3H\rho-\lambda\rho .
\label{eq:closed_rho}
\end{eqnarray}
This system should be understood as the reduced field equations
supplemented by the minimal higher-dimensional conservation closure. The
coupling between matter and geometry is encoded in the term
$\lambda\rho$, which describes the dilution of matter by the internal
volume modulus.

\subsection{Dimensionless cosmological functions in the coupled case}

For comparison with standard cosmology, it is convenient to introduce the
dimensionless Hubble function
\begin{eqnarray}
E_U(z)
\equiv
\frac{H(z)}{H_0},
\label{eq:EU_def}
\end{eqnarray}
where $H_0$ is the present value of the Hubble parameter. We use $z$ only
to denote the cosmological redshift, defined by
\begin{eqnarray}
1+z=\frac{a_0}{a},
\qquad
N\equiv \ln a.
\label{eq:z_N_def}
\end{eqnarray}
Therefore,
\begin{eqnarray}
\frac{d\ln H}{dN}
=
\frac{\dot H}{H^2}.
\label{eq:dlnH_dN}
\end{eqnarray}
For the coupled $d=1$ system, Eq.~(\ref{eq:closed_H}) gives
\begin{eqnarray}
\frac{d\ln H}{dN}
=
-1
-
\frac{8\pi G\rho}{3H^2}
+
\frac{\lambda}{H}.
\label{eq:dlnH_dN_d1}
\end{eqnarray}
We now define the Hubble-normalized matter variable and the dimensionless
geometric coupling variable as
\begin{eqnarray}
x\equiv \Omega_m
=
\frac{8\pi G\rho}{3H^2},
\qquad
y\equiv \frac{\lambda}{H}.
\label{eq:xy_def}
\end{eqnarray}
With this notation,
\begin{eqnarray}
\frac{H'}{H}
=
y-x-1,
\label{eq:Hprime_reduced}
\end{eqnarray}
where the prime denotes differentiation with respect to $N=\ln a$.

If the general closure Eq.~(\ref{eq:general_Q_closure}) is retained in
the pressureless case, it is useful to introduce the dimensionless exchange
function
\begin{eqnarray}
{\cal Q}
\equiv
\frac{8\pi G Q}{3H^3}.
\label{eq:calQ_def}
\end{eqnarray}
The reduced system then takes the form
\begin{eqnarray}
x'
=
2x^2-3xy-x+{\cal Q},
\label{eq:xprime_Q}
\end{eqnarray}
\begin{eqnarray}
y'
=
2-x+xy-2y^2.
\label{eq:yprime_Q}
\end{eqnarray}
The minimal closure adopted in this work corresponds to ${\cal Q}=0$.
Hence the reduced autonomous system used below is
\begin{eqnarray}
x'
=
2x^2-3xy-x,
\label{eq:xprime_reduced}
\end{eqnarray}
\begin{eqnarray}
y'
=
2-x+xy-2y^2.
\label{eq:yprime_reduced}
\end{eqnarray}
Equations~(\ref{eq:xprime_reduced}) and
(\ref{eq:yprime_reduced}) define the reduced matter--geometry phase
portrait, while Eq.~(\ref{eq:Hprime_reduced}) determines the
evolution of the Hubble variable.

It is important to distinguish two possible normalizations. The variable
$x=\Omega_m$ defined in Eq.~(\ref{eq:xy_def}) is normalized by the
instantaneous Hubble scale $H^2(z)$. Therefore, if the total effective
budget is normalized by $H^2(z)$, the geometric contribution is defined by
\begin{eqnarray}
\Omega_{\rm geo}^{(H)}(z)
\equiv
1-x(z),
\label{eq:Omega_geo_H}
\end{eqnarray}
so that
\begin{eqnarray}
1
=
x(z)+\Omega_{\rm geo}^{(H)}(z).
\label{eq:Omega_sum_H}
\end{eqnarray}
On the other hand, if the effective expansion history is written in the
usual form normalized by $H_0^2$, one should introduce
\begin{eqnarray}
\widehat\Omega_m(z)
\equiv
\frac{8\pi G\rho(z)}{3H_0^2}
=
E_U^2(z)x(z),
\label{eq:Omega_hat_m}
\end{eqnarray}
and define
\begin{eqnarray}
\widehat\Omega_{\rm geo}(z)
\equiv
E_U^2(z)-\widehat\Omega_m(z)
=
E_U^2(z)\left[1-x(z)\right].
\label{eq:Omega_hat_geo}
\end{eqnarray}
With this $H_0$-normalized convention, the effective decomposition reads
\begin{eqnarray}
E_U^2(z)
=
\widehat\Omega_m(z)
+
\widehat\Omega_{\rm geo}(z).
\label{eq:E_decomposition}
\end{eqnarray}

In the present minimally closed framework, the matter density parameter is
not expected to scale as the standard dust expression
$\Omega_{m0}(1+z)^3/E_U^2(z)$, because the matter evolution is modified by
the interaction with the internal-volume sector. Accordingly, the
functions $E_U(z)$, $x(z)$, $\Omega_{\rm geo}^{(H)}(z)$, and
$\widehat\Omega_{\rm geo}(z)$ must be reconstructed directly from the
closed dynamical system.

Another useful quantity is the deceleration parameter,
\begin{eqnarray}
q
\equiv
-1-\frac{\dot H}{H^2}.
\label{eq:q_def}
\end{eqnarray}
Using Eq.~(\ref{eq:Hprime_reduced}), one obtains
\begin{eqnarray}
q=x-y.
\label{eq:q_xy}
\end{eqnarray}
This expression shows explicitly that the effective deceleration is
controlled by the competition between the Hubble-normalized matter
content, $x=\Omega_m$, and the geometric coupling variable, $y=\lambda/H$.

Therefore, in the minimally closed coupled case, the cosmological
evolution is characterized by the functions $E_U(z)$, $x(z)$, $y(z)$,
$\Omega_{\rm geo}^{(H)}(z)$, $\widehat\Omega_{\rm geo}(z)$, and $q(z)$,
all reconstructed directly from the autonomous system. The main difference
with respect to the standard uncoupled dust case is that the matter
density does not evolve only with the external volume $a^3$, but also with
the internal-volume modulus $\psi$.

\section{Dynamical system, equilibrium points and Poincar\'e directions}

In this section we analyze the dynamical structure of the cosmological
systems obtained in Sec.~III. Our aim is not only to determine the finite
equilibrium configurations and the asymptotic directions that govern the
global behavior of the phase space, but also to identify which features of
the flow arise specifically from the unimodular higher-dimensional sector.
For this reason, the critical points and Poincar'e directions will be
interpreted not merely as mathematical properties of the reduced dynamical
system, but as physical signatures of the internal-volume degree of freedom.
In particular, they provide a direct way to assess how the extra-dimensional
unimodular geometry modifies the late-time cosmological evolution and the
global organization of the phase space.

This point is important because, in the present framework, the dimensional
reduction does not simply add an auxiliary scalar variable to an otherwise
standard cosmology. Instead, it reorganizes the effective dynamics itself.
In the vacuum sector, the reduced system admits a continuous finite
equilibrium set. In the matter sector, after the reduced Bianchi relation
has been interpreted as an integrability condition, the dynamics must be
closed by a physical prescription for the matter sector. In this work we
adopt the minimal closure \(Q=0\), introduced in Sec.~III, for which the
matter density is diluted by both the external volume \(a^3\) and the
internal-volume modulus \(\psi\). Under this closure the coupled
matter--geometry system is autonomous and its reduced phase space is
organized by isolated critical points.

We consider separately the vacuum system and the minimally closed system
with matter. In each case, we first examine the finite critical structure
and then the behavior at infinity through the corresponding Poincar\'e
directions and compactified representation.

\subsection{Equilibrium points for the vacuum system}

We begin with the vacuum system derived in Sec.~III,
Eqs.~(\ref{eq:Hdot_vac}) and (\ref{eq:lambdadot_vac}). Since the vector
field is quadratic in \((H,\lambda)\), the analysis of the equilibrium
structure can be performed directly in the phase plane.

\subsection*{Finite equilibrium points}

The finite equilibrium points are obtained by imposing
\begin{eqnarray}
\dot H=0,
\qquad
\dot\lambda=0.
\label{eq:vaccrit_cond}
\end{eqnarray}
Using Eqs.~(\ref{eq:Hdot_vac}) and (\ref{eq:lambdadot_vac}), this gives
\begin{eqnarray}
3d^2H^2+(1-d)\lambda^2+d(d-4)H\lambda=0,
\label{eq:vac_alg1}
\\
6dH^2-3\lambda^2+3(d-2)H\lambda=0.
\label{eq:vac_alg2}
\end{eqnarray}
The origin,
\begin{eqnarray}
(H,\lambda)=(0,0),
\label{eq:vac_origin}
\end{eqnarray}
is an immediate solution. Since the system is purely quadratic, the
Jacobian matrix vanishes at this point, and therefore the origin is
non-hyperbolic.

For nontrivial equilibrium points with \(H\neq 0\), it is convenient to
introduce the ratio
\begin{eqnarray}
y\equiv \frac{\lambda}{H}.
\label{eq:yratio_vac}
\end{eqnarray}
Here and in what follows, the symbol \(z\) is reserved for the
cosmological redshift, while \(y\) denotes the ratio between the
internal-volume rate and the Hubble rate. Dividing
Eq.~(\ref{eq:vac_alg2}) by \(3H^2\), one obtains
\begin{eqnarray}
y^2-(d-2)y-2d=0,
\label{eq:vac_y_poly}
\end{eqnarray}
whose roots are
\begin{eqnarray}
y=d,
\qquad
y=-2.
\label{eq:vac_y_roots}
\end{eqnarray}
Substituting these values into Eq.~(\ref{eq:vac_alg1}), one finds that
only \(y=d\) is admissible for \(d>0\). Hence, the nontrivial finite
equilibrium set is the straight line
\begin{eqnarray}
\lambda=dH,
\qquad
H\in\mathbb{R}.
\label{eq:vac_line}
\end{eqnarray}

Thus, besides the degenerate origin, the vacuum system admits a
one-parameter family of equilibrium points. This result contrasts sharply
with the situation in standard four-dimensional General Relativity. In the
absence of matter and a cosmological constant, the FLRW equations of
General Relativity admit only the Minkowski solution, corresponding to the
isolated critical point $(H=0)$ The phase-space structure is therefore
organized around a single finite equilibrium configuration. In the present
unimodular higher-dimensional model, by contrast, the reduced dynamics
admits an entire equilibrium line, $(\lambda=dH)$. This implies that the
extra-dimensional scalar sector does not merely perturb the flow around an
isolated state, but generates a continuous family of balanced
configurations in which the Hubble parameter and the internal-volume degree
of freedom evolve in a correlated manner. The existence of the equilibrium
line  $(\lambda=dH) $ is therefore a direct dynamical consequence of the
higher-dimensional unimodular sector and represents one of the most
distinctive features of the model.

\subsection*{Poincar\'e directions and compactified vacuum flow}

We now examine the behavior of the vacuum system at infinity. In this
sector the original phase space is the \((H,\lambda)\) plane, where
\(\lambda=\dot\psi/\psi\). Therefore, the coordinates used in the
Poincar\'e compactification should not be confused with the matter
variable \(x=\Omega_m\), which is introduced only in the coupled matter
system.

A convenient compactification of the vacuum plane is
\begin{eqnarray}
X_{\rm P}
=
\frac{H}{\sqrt{1+H^2+\lambda^2}},
\qquad
Y_{\rm P}
=
\frac{\lambda}{\sqrt{1+H^2+\lambda^2}},
\qquad
Z_{\rm P}
=
\frac{1}{\sqrt{1+H^2+\lambda^2}}.
\label{eq:vac_poincare_coordinates}
\end{eqnarray}
These variables satisfy
\begin{eqnarray}
X_{\rm P}^2+Y_{\rm P}^2+Z_{\rm P}^2=1,
\qquad
Z_{\rm P}\geq 0.
\label{eq:poincare_sphere_vac}
\end{eqnarray}
Thus the finite phase plane is mapped into the upper Poincar\'e
hemisphere, while the equator \(Z_{\rm P}=0\) represents infinity. The
projection onto the \((X_{\rm P},Y_{\rm P})\) plane gives the usual
Poincar\'e disk. Since both \(H\) and \(\lambda\) are divided by the same
radial factor, one has
\begin{eqnarray}
\frac{Y_{\rm P}}{X_{\rm P}}
=
\frac{\lambda}{H}.
\label{eq:vac_poincare_ratio}
\end{eqnarray}
Thus, each point on the Poincar\'e boundary represents an asymptotic
direction in the original \((H,\lambda)\) plane.

Here it is also useful to introduce the projective variable analogously to equation (\ref{eq:yratio_vac})
together with the e-fold variable \(N=\ln a\). The variable \(y\) labels
the slope of a direction in the \((H,\lambda)\) plane, or equivalently the
ratio \(Y_{\rm P}/X_{\rm P}\) on the compactified boundary. It should not
be identified with the vertical coordinate of the disk itself; rather, it
is a projective coordinate that selects an asymptotic direction.

Using the vacuum autonomous system derived in Sec.~III, the angular
dynamics can be written as
\begin{eqnarray}
y'
=
\frac{dy}{dN}
=
\frac{1}{(d+2)}
h(y),
\label{eq:vac_yprime}
\end{eqnarray}
where, up to the irrelevant positive factor \(d(d+2)\),
\begin{eqnarray}
h(y)
=
\frac{1-d}{d}y^3+(d-7)y^2+6(d-1)y+6d.
\label{eq:vac_hy}
\end{eqnarray}
The Poincar\'e directions at infinity are therefore determined by the
roots of \(h(y)=0\). This polynomial factorizes as
\begin{eqnarray}
h(y)
=
(y-d)
\left[
\frac{1-d}{d}y^2-6y-6
\right].
\label{eq:vac_hy_fact}
\end{eqnarray}
Hence one always obtains the direction
\begin{eqnarray}
y_0=d,
\label{eq:vac_y0}
\end{eqnarray}
which corresponds to the same ratio that defines the finite equilibrium
line \(\lambda=dH\). For \(d\neq 1\), the other two asymptotic directions
are
\begin{eqnarray}
y_{\pm}
=
\frac{-3d\pm\sqrt{3d(d+2)}}{d-1}.
\label{eq:vac_infdirs}
\end{eqnarray}

The case \(d=1\) must be treated separately, since the expression above
for \(y_{\pm}\) becomes singular. In this case the polynomial reduces to
\begin{eqnarray}
h(y)=6(1-y^2),
\label{eq:vac_hy_d1}
\end{eqnarray}
and the Poincar\'e directions are
\begin{eqnarray}
y=1,
\qquad
y=-1.
\label{eq:vac_infdirs_d1}
\end{eqnarray}
The direction \(y=1\) is the compactified image of the vacuum equilibrium
line \(\lambda=H\) for \(d=1\), while \(y=-1\) represents the opposite
asymptotic sector of the flow.

These directions determine the asymptotic organization of the vacuum
phase portrait. In particular, the direction \(y_0=d\) is not accidental:
it is the same ratio \(\lambda/H\) that characterizes the finite
equilibrium line \(\lambda=dH\). In the compactified disk this direction
is represented by
\begin{eqnarray}
\frac{Y_{\rm P}}{X_{\rm P}}=d,
\label{eq:vac_compact_line}
\end{eqnarray}
or equivalently \(Y_{\rm P}=dX_{\rm P}\). Therefore, the same balance
between the Hubble rate and the internal-volume variable that appears at
finite points also controls one of the preferred directions at infinity.

This provides a global interpretation of the reduced vacuum flow. The
extra-dimensional sector is visible not only through the finite critical
line, but also through the angular structure of the compactified phase
space. Consequently, the difference with respect to the relativistic
reference case is global in phase-space terms, and not merely local around
finite critical configurations.

\subsection{Equilibrium points for the system with matter}

We now consider the minimally closed cosmological system with matter
obtained in Sec.~III. In the corrected formulation, the reduced Bianchi
relation is not used as an independent equation for the matter density.
Instead, it is interpreted as an integrability condition. The matter
evolution is then specified by the exchange law
\begin{eqnarray}
\dot\rho+3H(\rho+p)+\lambda\rho=Q.
\label{eq:Q_recalled_IV}
\end{eqnarray}
The present phase-space analysis is performed for the minimal closure
\(Q=0\), for which pressureless matter satisfies
\begin{eqnarray}
\dot\rho+3H\rho+\lambda\rho=0.
\label{eq:minimal_closure_IV}
\end{eqnarray}
This choice corresponds to dilution by the external volume and by the
internal-volume modulus, \(\rho a^3\psi=\mathrm{constant}\), and implies
\(\dot R=0\) through the reduced Bianchi relation. Other choices of
\(Q\) would define different coupled systems and would require a separate
critical-point and compactification analysis.

For \(d=1\), the minimally closed system is governed by
Eqs.~(\ref{eq:closed_H})--(\ref{eq:closed_rho}). At the level of the full
variables \((H,\lambda,\rho)\), the system is three-dimensional. However,
for the phase-space analysis it is convenient to introduce the reduced
variables
\begin{eqnarray}
x\equiv \Omega_m
=
\frac{8\pi G\rho}{3H^2},
\qquad
y\equiv \frac{\lambda}{H},
\label{eq:reduced_variables_IV}
\end{eqnarray}
which define the reduced phase plane. In these variables, and using the
minimal closure \(Q=0\), one obtains
\begin{eqnarray}
x'
=
2x^2-3xy-x,
\label{eq:xprime_IV}
\\
y'
=
2-x+xy-2y^2,
\label{eq:yprime_IV}
\end{eqnarray}
where the prime denotes differentiation with respect to the e-fold
variable \(N=\ln a\). The quantity
\begin{eqnarray}
S\equiv \frac{H'}{H}=y-x-1
\label{eq:S_IV}
\end{eqnarray}
controls the evolution of the Hubble variable in the augmented
representation \((x,y,H)\), or equivalently \((x,y,E)\) with
\(E=H/H_0\).

It is worth stressing that
Eqs.~(\ref{eq:xprime_IV})--(\ref{eq:yprime_IV}) are the reduced
dynamical system associated with the minimally closed full system.
Therefore, the phase portrait in the \((x,y)\) plane and its
version in \((x,y,H)\) provide a faithful representation of the reduced
dynamics, although they should not be confused with the full phase space
\((H,\lambda,\rho)\) itself.

\subsection*{Finite equilibrium points}

The finite equilibrium points of the reduced system satisfy
\begin{eqnarray}
x'=0,
\qquad
y'=0.
\label{eq:crit_cond_matter_new}
\end{eqnarray}
From Eq.~(\ref{eq:xprime_IV}), one has
\begin{eqnarray}
x(2x-3y-1)=0,
\label{eq:xcrit_factor}
\end{eqnarray}
so that either
\begin{eqnarray}
x=0,
\label{eq:x0_branch}
\end{eqnarray}
or
\begin{eqnarray}
2x-3y-1=0.
\label{eq:x_nonzero_branch}
\end{eqnarray}

If \(x=0\), Eq.~(\ref{eq:yprime_IV}) reduces to
\begin{eqnarray}
2-2y^2=0,
\end{eqnarray}
whose solutions are
\begin{eqnarray}
y=\pm 1.
\end{eqnarray}
This yields the two critical points
\begin{eqnarray}
C_+=(0,1),
\qquad
C_-=(0,-1).
\label{eq:Cpm_points}
\end{eqnarray}

If instead Eq.~(\ref{eq:x_nonzero_branch}) holds, then
\begin{eqnarray}
x=\frac{3y+1}{2}.
\label{eq:x_of_y}
\end{eqnarray}
Substituting this into Eq.~(\ref{eq:yprime_IV}), one finds
\begin{eqnarray}
2-\frac{3y+1}{2}
+y\left(\frac{3y+1}{2}\right)
-2y^2=0,
\end{eqnarray}
which simplifies to
\begin{eqnarray}
\frac{3-2y-y^2}{2}=0.
\label{eq:y_aux_poly}
\end{eqnarray}
The solutions are
\begin{eqnarray}
y=1,
\qquad
y=-3.
\end{eqnarray}
The second solution gives \(x=-4\) and is outside the physical domain
\(x=\Omega_m\geq 0\). Hence the admissible nontrivial solution is
\begin{eqnarray}
y=1,
\qquad
x=2.
\end{eqnarray}
Therefore, the third finite critical point is
\begin{eqnarray}
M=\left(2,1\right).
\label{eq:M_point}
\end{eqnarray}

Thus, within the physical domain \(x\geq 0\), the minimally closed
coupled matter--geometry system admits three finite critical points in
the reduced phase plane,
\begin{eqnarray}
C_+=(0,1),
\qquad
C_-=(0,-1),
\qquad
M=\left(2,1\right).
\label{eq:all_critical_points}
\end{eqnarray}

The linear classification follows from the Jacobian matrix of
Eqs.~(\ref{eq:xprime_IV})--(\ref{eq:yprime_IV}),
\begin{eqnarray}
J(x,y)
=
\begin{pmatrix}
4x-3y-1 & -3x \\
y-1 & x-4y
\end{pmatrix}.
\label{eq:jacobian_matter}
\end{eqnarray}
At the three finite critical points one obtains
\begin{eqnarray}
C_+:\quad
\{-4,-4\},
\qquad
C_-:\quad
\{2,4\},
\qquad
M:\quad
\{4,-2\}.
\label{eq:eigenvalues_matter}
\end{eqnarray}
Therefore,
\begin{eqnarray}
C_+:\ \mbox{attractor},
\qquad
C_-:\ \mbox{repeller},
\qquad
M:\ \mbox{saddle}.
\label{eq:critical_classification_new}
\end{eqnarray}

This structure should be contrasted with the vacuum case. There, the
reduced unimodular dynamics is organized by the continuous equilibrium
line \(\lambda=dH\). In the minimally closed coupled matter system this
degeneracy is lifted and replaced by isolated critical configurations.
In other words, once the matter sector is coupled to the internal-volume
modulus through the closure \(Q=0\), the phase-space organization becomes
more selective: the flow no longer admits a one-parameter family of finite
balanced states, but instead distinguishes an attractive branch, a
repulsive branch, and an intermediate saddle configuration.

More specifically, \(C_+\) represents the late-time attractive sector of
the reduced dynamics, \(C_-\) corresponds to the repulsive branch, and
\(M\) plays the role of a transition configuration separating different
flow channels. From the dynamical point of view, this means that the
matter--geometry coupling does not erase the imprint of the
extra-dimensional sector, but reorganizes it. The qualitative novelty is
therefore not just the existence of critical points, but the replacement
of the vacuum critical line by a finite set of isolated configurations
with different stability properties.

\subsection*{Poincar\'e directions and compactified matter flow}

We now analyze the asymptotic behavior of the minimally closed
matter-coupled system. The Poincar\'e construction used here is analogous
to the one introduced in the vacuum sector, but it is applied to a
different phase plane. In the vacuum case, the compactification was
performed in the original $(H,\lambda)$ plane, where the projective
variable $\lambda/H$ only labelled directions at infinity. In the
matter-coupled case, by contrast, the relevant autonomous system is the
reduced normalized system written in terms of
\begin{equation}
x=\Omega_m=\frac{8\pi G\rho}{3H^2},
\qquad
y=\frac{\lambda}{H}.
\end{equation}
Therefore, in this sector $x$ and $y$ are not Poincar\'e coordinates.
They are the physical reduced variables of the matter--geometry system:
$x$ measures the Hubble-normalized matter density, while $y$ measures the
internal-volume rate normalized by the Hubble expansion.

Accordingly, we use the same compactification prescription defined for
the vacuum analysis, but now applied to the reduced $(x,y)$ plane. The
resulting Poincar\'e disk, or equivalently the upper Poincar\'e hemisphere,
represents the global organization of the normalized matter--geometry
flow. The compactified coordinates themselves should therefore be regarded
only as graphical coordinates on the disk or hemisphere, and not as new
cosmological density parameters.

This distinction is important because the original matter-coupled system
is three-dimensional in the variables $(H,\lambda,\rho)$. The reduction to
$(x,y)$ removes the overall Hubble scale and keeps only the normalized
matter content and the normalized internal-volume rate. The Hubble
variable is then reconstructed separately from
\begin{equation}
\frac{E'}{E}=y-x-1,
\qquad
E=\frac{H}{H_0}.
\end{equation}
Thus, the compactified $(x,y)$ portrait describes the global structure of
the reduced autonomous flow. The additional variable $E=H/H_0$ does not
define a new independent phase-space dynamics; it only tracks how the
Hubble scale changes along the trajectories determined by $(x,y)$.

For the minimal closure $Q=0$, the reduced autonomous system is
\begin{equation}
x'=2x^2-3xy-x,
\qquad
y'=2-x+xy-2y^2 .
\end{equation}
The directions at infinity are determined only by the highest-degree terms
of this vector field. Hence we retain the homogeneous quadratic part,
\begin{equation}
x'_{\rm hom}=2x^2-3xy,
\qquad
y'_{\rm hom}=xy-2y^2 .
\end{equation}
On the Poincar\'e boundary, an asymptotic direction is selected when the
leading vector field is parallel to the radial direction in the reduced
$(x,y)$ plane. This gives the projective condition
\begin{equation}
y x'_{\rm hom}-x y'_{\rm hom}=0 .
\end{equation}
Substituting the homogeneous terms, one obtains
\begin{equation}
y(2x^2-3xy)-x(xy-2y^2)=xy(x-y)=0 .
\end{equation}
Therefore, the asymptotic directions of the reduced coupled system are
\begin{equation}
x=0,
\qquad
y=0,
\qquad
y=x .
\end{equation}

These three directions form the asymptotic skeleton of the compactified
matter--geometry flow. The direction $x=0$ corresponds to the matter-free
boundary of the reduced plane, the direction $y=0$ corresponds to a
vanishing Hubble-normalized internal-volume rate, and the diagonal
direction $y=x$ represents an asymptotic balance between the matter
contribution and the geometric scalar contribution. Thus, although the
matter coupling changes the finite critical structure by replacing the
vacuum equilibrium line with the isolated points $C_+$, $C_-$, and $M$,
the compactified reduced flow remains organized by a small number of
well-defined directions at infinity.

\section{Some Numerical Results}

In this section we present numerical illustrations of the main dynamical
features derived in the previous sections. The goal is not to perform an
observational fit, but to show how the analytical phase-space structure
appears in representative cosmological trajectories and compactified
flows. Throughout the main text we focus on the five-dimensional case
$d=1$, for which the coupled matter--geometry system was derived
explicitly. In the reduced variables given by equations 
\begin{eqnarray}
x\equiv \Omega_m=\frac{8\pi G\rho}{3H^2},
\qquad
y\equiv \frac{\lambda}{H},
\end{eqnarray}
the reduced system is governed by Eqs.~(70)-(71), together with
Eq.~(66) for the evolution of the Hubble variable.

\subsection*{A. Background evolution relative to the $\Lambda$CDM reference}

\begin{figure}[htbp]
    \centering
    \includegraphics[width=0.440\textwidth]{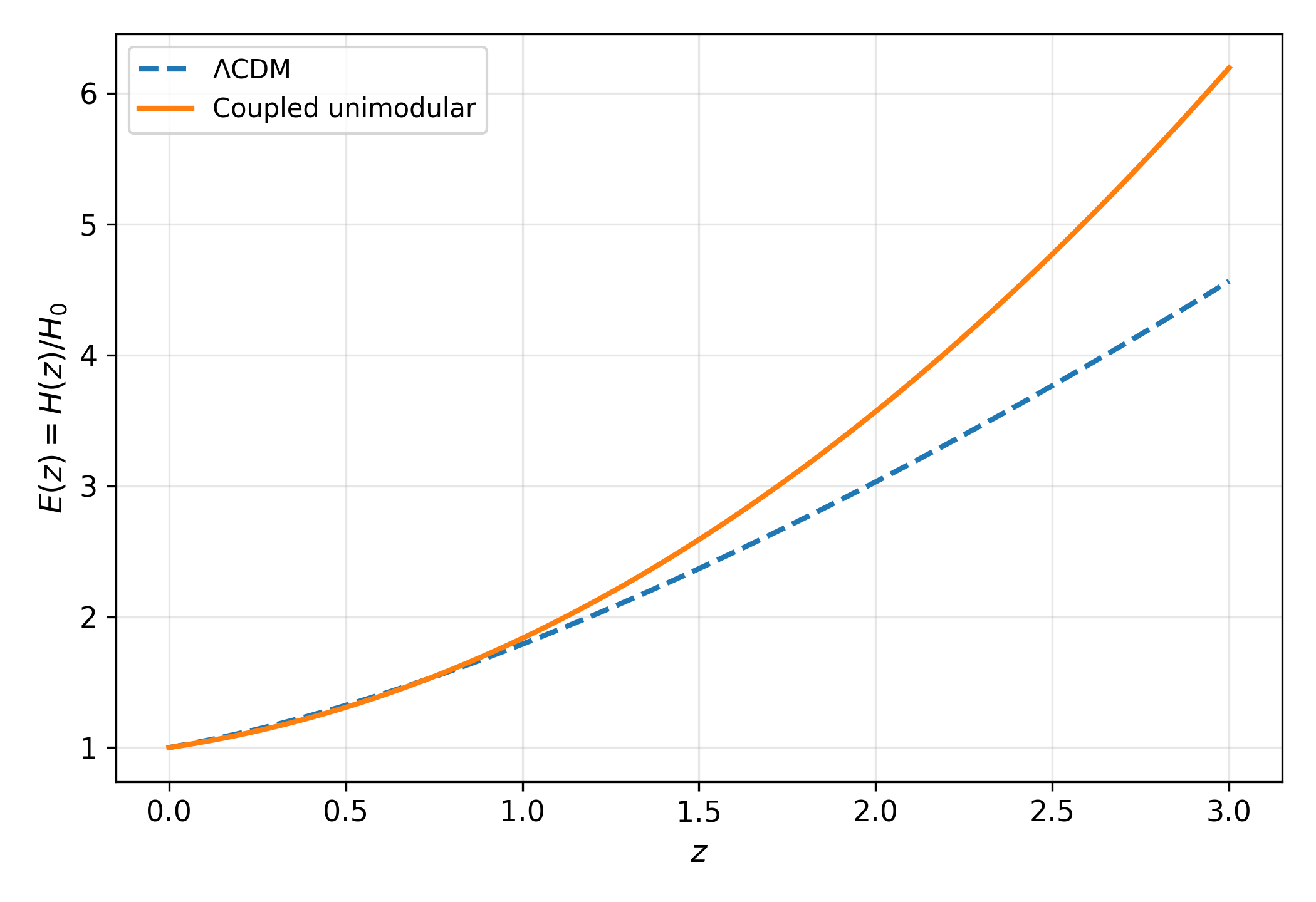}
    \includegraphics[width=0.440\textwidth]{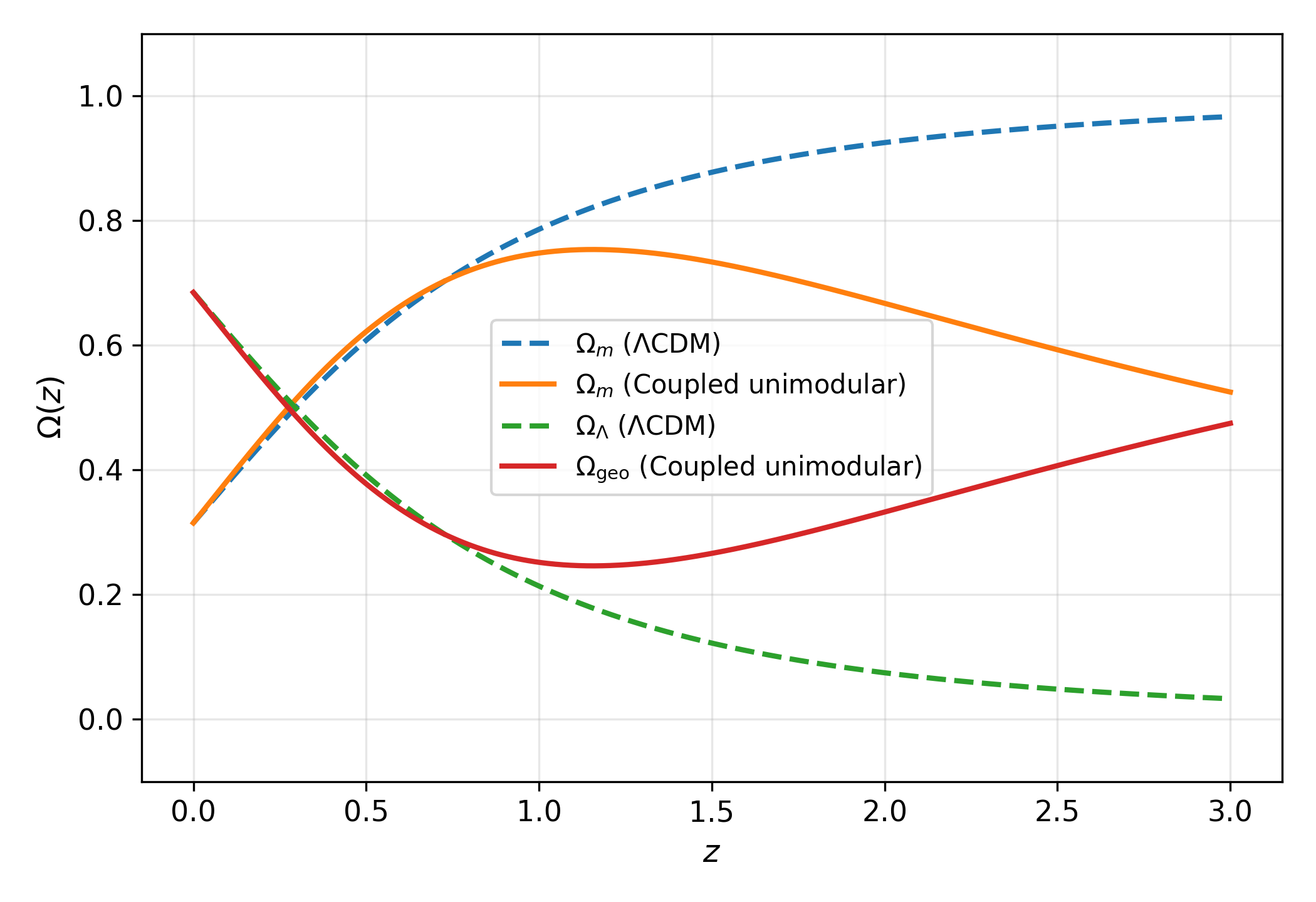}
    \includegraphics[width=0.440\textwidth]{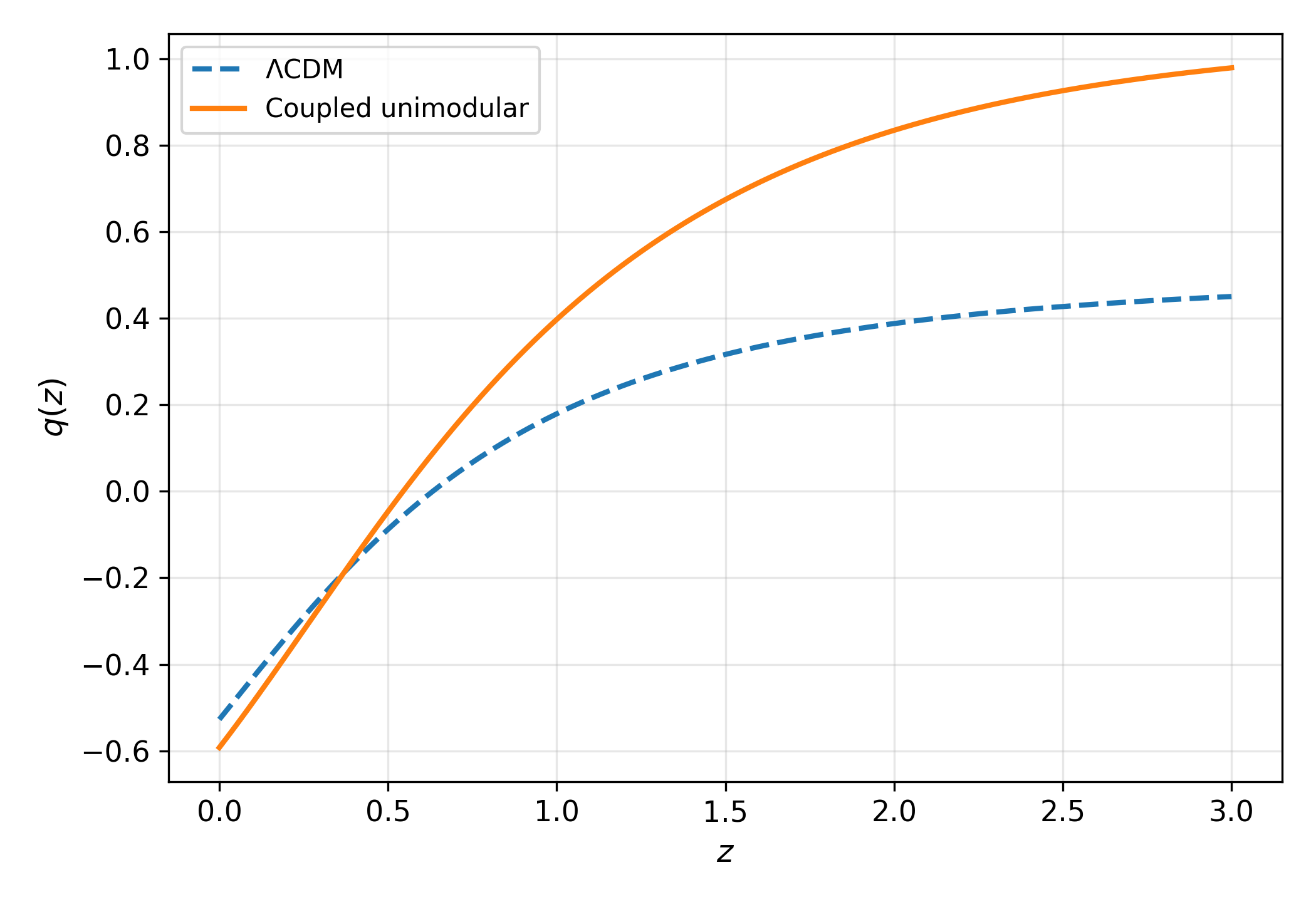}
    \includegraphics[width=0.440\textwidth]{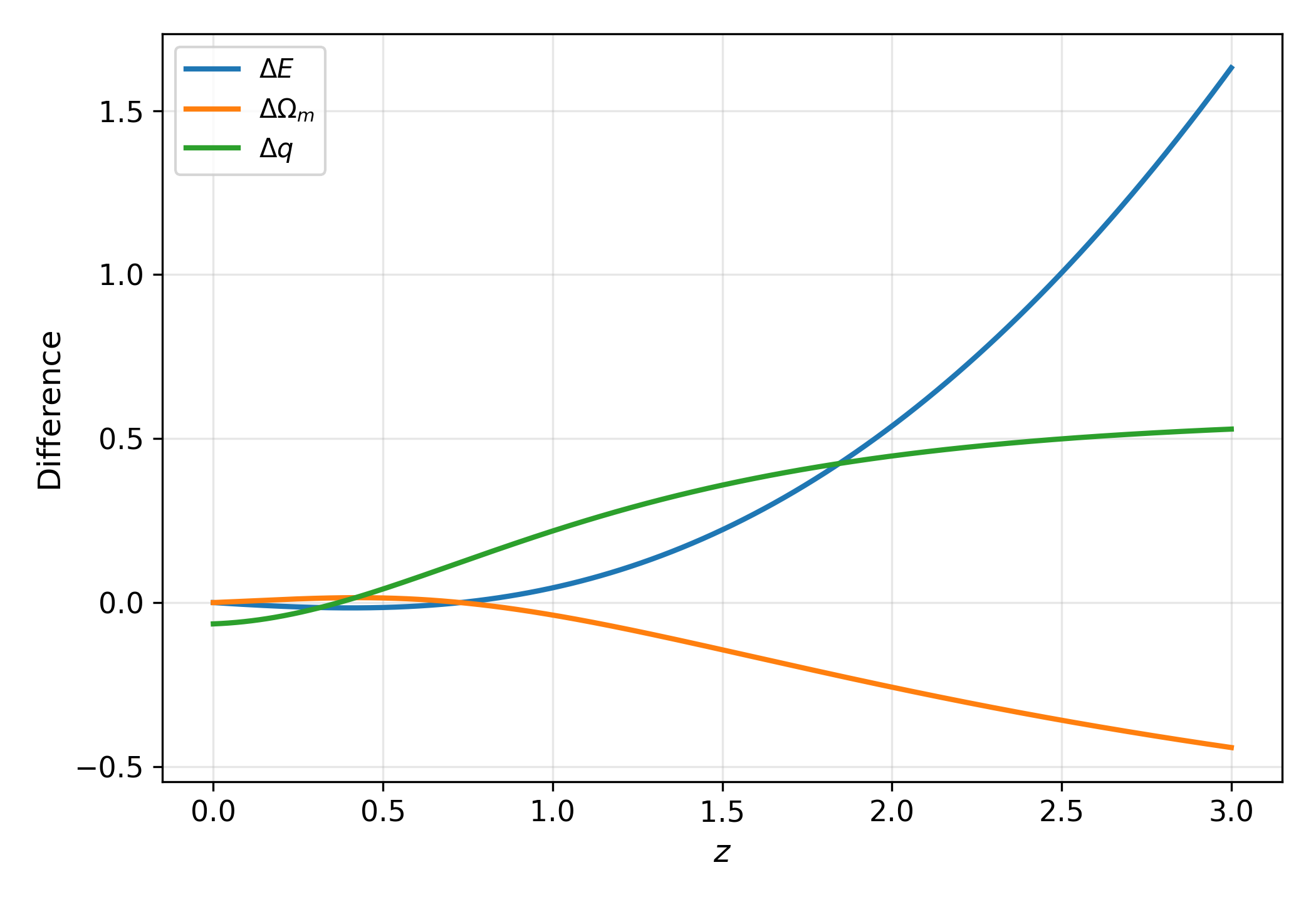}
   \caption{
Comparison between the coupled unimodular model and the $\Lambda$CDM
reference for the main background observables in the five-dimensional
case $d=1$. The numerical integration uses
$\Omega_{m0}=0.315$, $\Omega_{\Lambda0}=0.685$, $E(0)=1$,
$x(0)=\Omega_{m0}$, and
$y(0)=y_{\rm match}(0)+\epsilon_y=0.9075$, with
$y_{\rm match}(0)=1-\Omega_{m0}/2=0.8425$ and
$\epsilon_y=0.065$. The panels show the dimensionless Hubble function
$E(z)=H(z)/H_0$, the $H^2(z)$-normalized density variables
$x(z)=\Omega_m(z)$ and $\Omega_{\rm geo}(z)=1-x(z)$, the deceleration
parameter $q(z)=x(z)-y(z)$, and the corresponding residuals with respect
to $\Lambda$CDM over the interval $0\leq z\leq 3$. The comparison is
intended as a reference benchmark: the coupled unimodular solution remains
close to $\Lambda$CDM near the present epoch, while deviations increase
towards higher redshift.
}
    \label{fig:background_compare}
\end{figure}

Figure~\ref{fig:background_compare} compares a representative coupled
unimodular solution with the $\Lambda$CDM reference at the level of the
background observables. The panels show the dimensionless Hubble function
$E(z)$, the $H^2(z)$-normalized density variables, the deceleration
parameter, and the corresponding residuals. In the coupled model these
quantities are reconstructed directly from the autonomous system, with
\begin{eqnarray}
\Omega_{\rm geo}(z)=1-x(z),
\qquad
q(z)=x(z)-y(z).
\end{eqnarray}

For the illustrative initial conditions used in the figure, the
unimodular solution remains close to the $\Lambda$CDM reference near the
present epoch, while deviations become more visible at higher redshift.
These deviations appear simultaneously in the expansion rate, in the
effective matter--geometry balance, and in the deceleration history. They
therefore reflect the dynamical influence of the internal-volume scalar
rather than a simple rescaling of the standard background evolution.

This comparison should not be interpreted as an observational constraint
on the model. A quantitative confrontation with data would require a
statistical fit of the free parameters and initial conditions, for
example using supernovae, cosmic chronometers, BAO, or other background
probes. Here $\Lambda$CDM is used only as a reference benchmark, in order
to display how the matter--geometry coupling modifies the background
evolution.

\subsection*{B. Evolution of the geometric coupling variable}

\begin{figure}[htbp]
    \centering
    \includegraphics[width=0.600\textwidth]{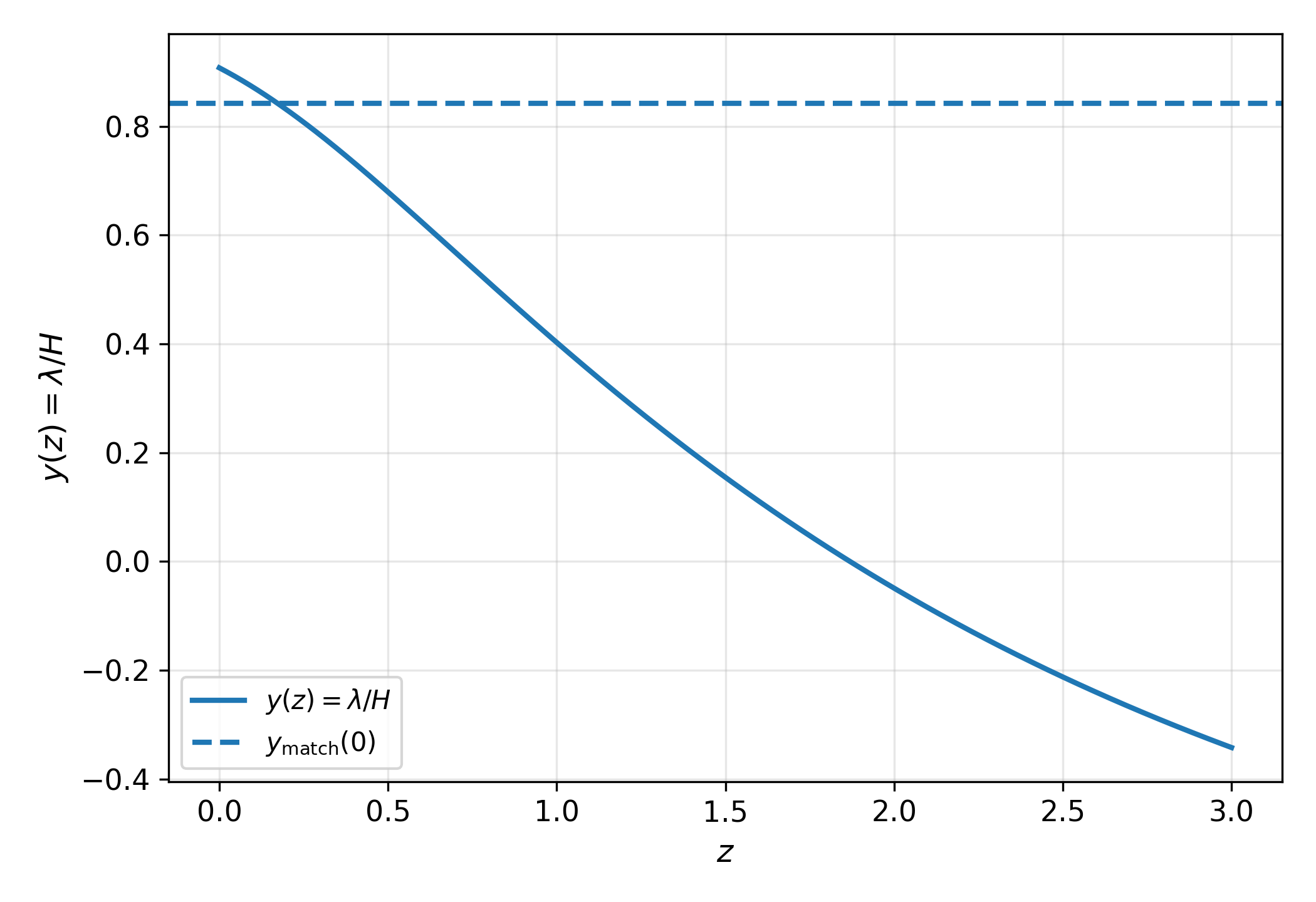}
    \caption{
Evolution of the geometric coupling variable
$y(z)=\lambda/H$ in the coupled unimodular model for $d=1$. The
    \label{fig:y_of_z}
integration uses $\Omega_{m0}=0.315$, $E(0)=1$,
$x(0)=\Omega_{m0}$, and
$y(0)=0.9075$, obtained from
$y_{\rm match}(0)=1-\Omega_{m0}/2=0.8425$ with
$\epsilon_y=0.065$. The horizontal dashed line indicates the
$\Lambda$CDM-matching value $y_{\rm match}(0)$. The monotonic decrease of
$y(z)$ over $0\leq z\leq 3$ shows that the internal-volume contribution
evolves nontrivially and controls the departure from the standard
background behavior.
}
    \label{fig:x_of_z}
\end{figure}

Figure~\ref{fig:y_of_z} shows the evolution of the dimensionless
geometric coupling variable
\begin{eqnarray}
y(z)=\frac{\lambda}{H}.
\end{eqnarray}
This quantity measures the rate of change of the internal-volume scalar
relative to the Hubble expansion and is therefore a direct diagnostic of
the extra-dimensional contribution.

The numerical solution shows that $y(z)$ evolves monotonically over the
redshift interval considered. This indicates that the geometric scalar
sector is not frozen and cannot be represented by a constant background
term. Instead, its evolution drives the progressive departure of the
coupled unimodular model from the $\Lambda$CDM reference. In this sense,
Fig.~\ref{fig:y_of_z} provides the most direct visualization of the
dynamical role played by the internal-volume degree of freedom in the
coupled system.

\subsection*{C. Vacuum phase portraits: comparison with the General Relativity}

\begin{figure}[htbp]
    \centering
    \includegraphics[width=0.400\textwidth]{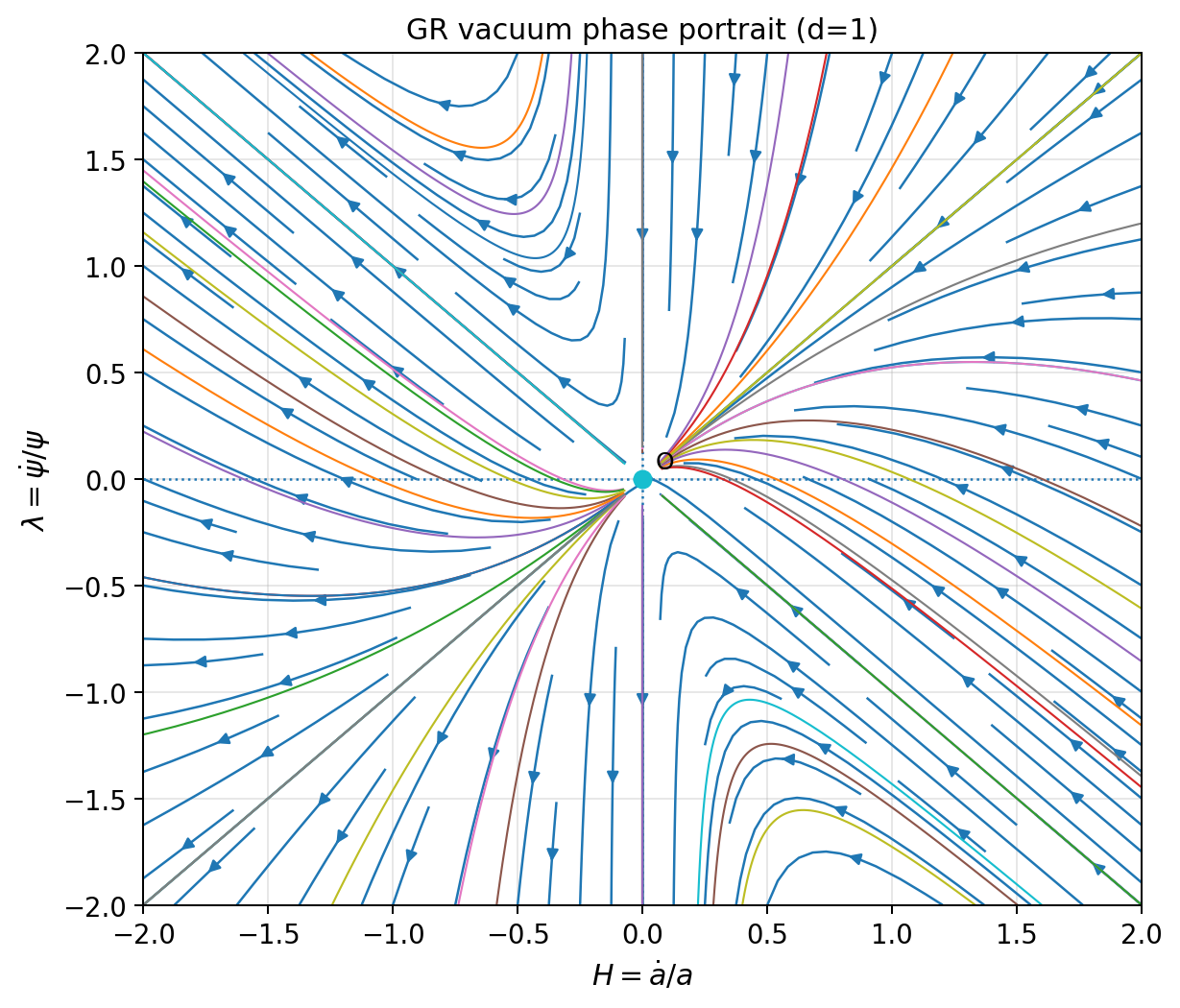}
    \includegraphics[width=0.400\textwidth]{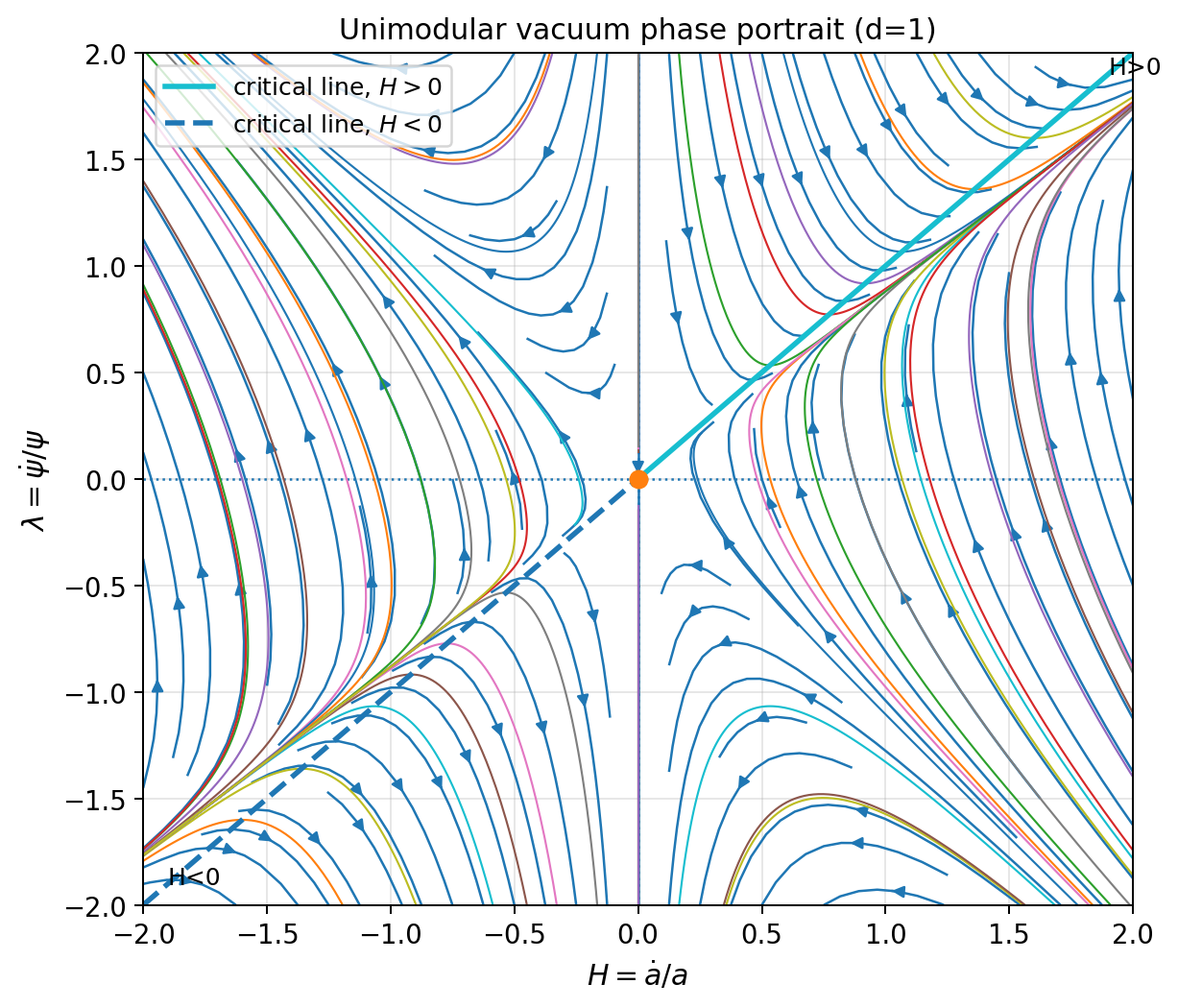}
    \caption{Vacuum phase portraits for general relativity and unimodular cosmology in the representative case $d=1$. While the General Relativity admits only the origin as a finite critical configuration, the unimodular system displays the critical line $\lambda=dH$, revealing the nontrivial dynamical role of the extra-dimensional sector already at the vacuum level.}
    \label{fig:vacuum_phase_portraits}
\end{figure}

Figure~\ref{fig:vacuum_phase_portraits} compares the vacuum phase
portraits of the general relativity and of the unimodular
reduced system for $d=1$. In the relativistic case, the finite phase
portrait is organized by the origin as the only finite critical
configuration. In the unimodular case, by contrast, the flow is organized
by the equilibrium line
\begin{eqnarray}
\lambda=dH.
\end{eqnarray}

This difference is not merely graphical. It shows that the dimensional
reduction of the unimodular theory produces a qualitatively distinct
vacuum dynamics. The Hubble variable and the internal-volume scalar can
remain dynamically balanced along an entire one-parameter family of
states. Thus, already in vacuum, the extra-dimensional sector changes the
finite phase-space organization of the reduced cosmological flow.

\subsection*{D. Global vacuum structure from Poincar'e compactification}

\begin{figure}[htbp]
    \centering
    \includegraphics[width=0.400\textwidth]{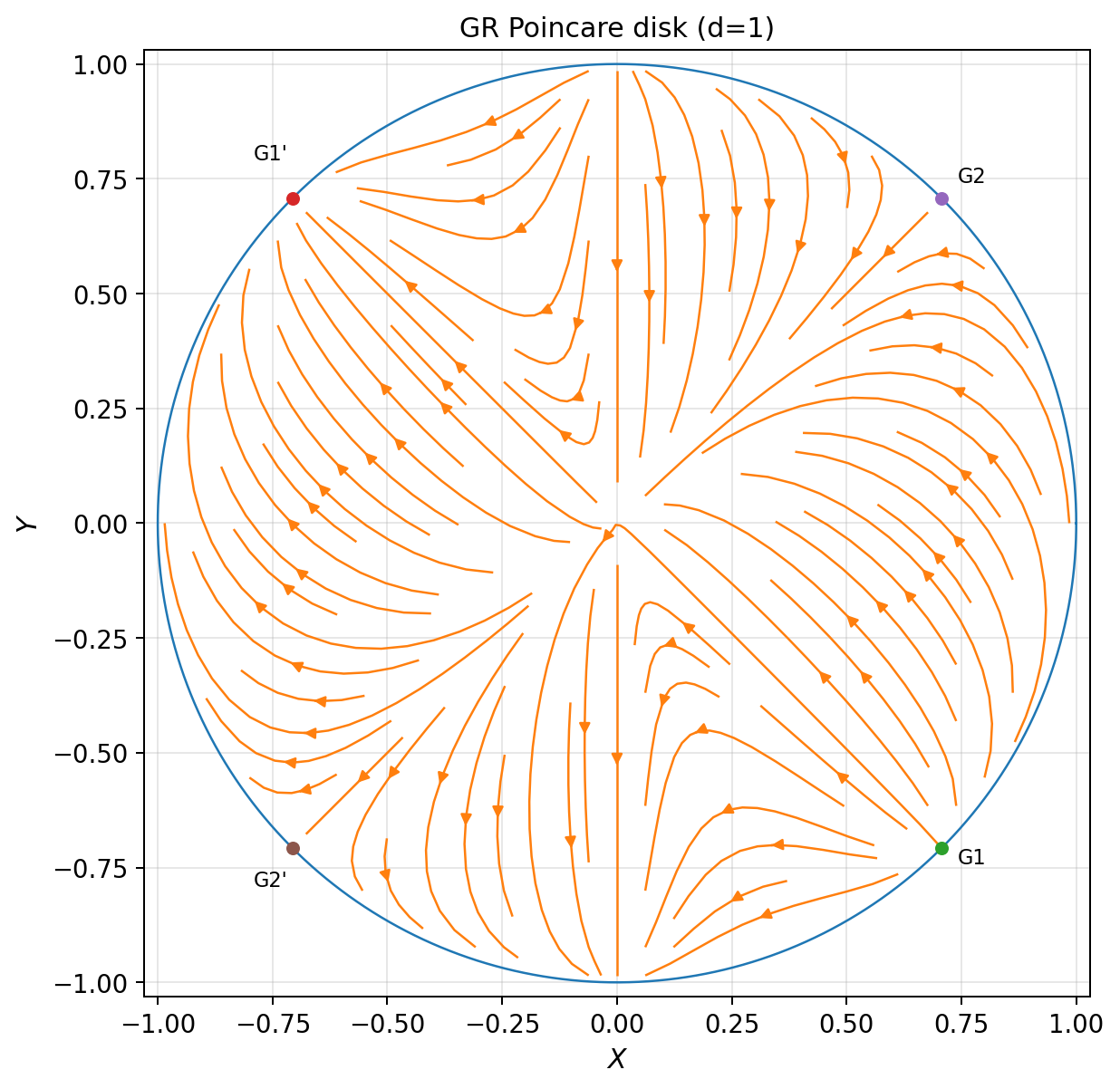}
    \includegraphics[width=0.400\textwidth]{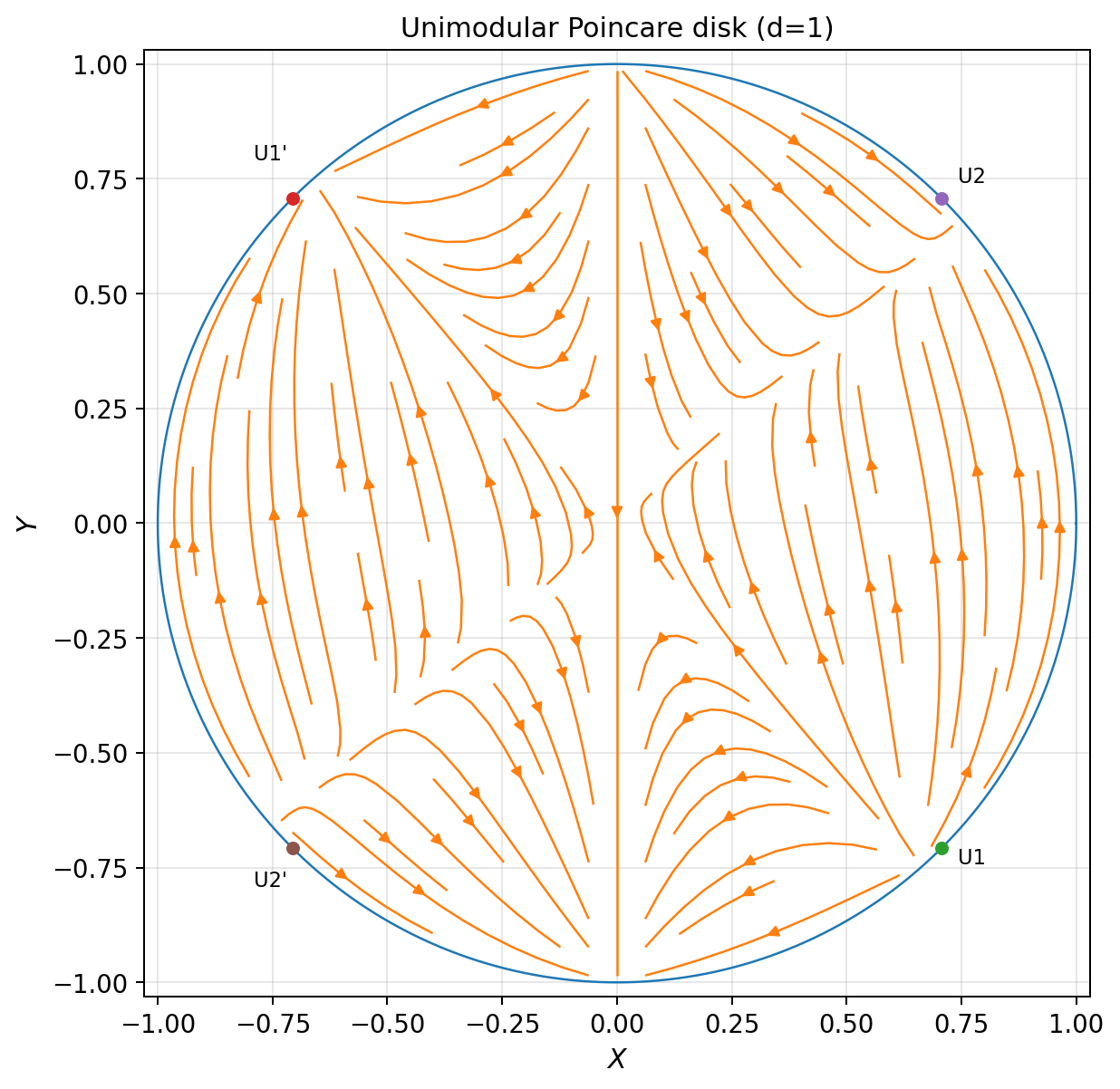}
    \includegraphics[width=0.400\textwidth]{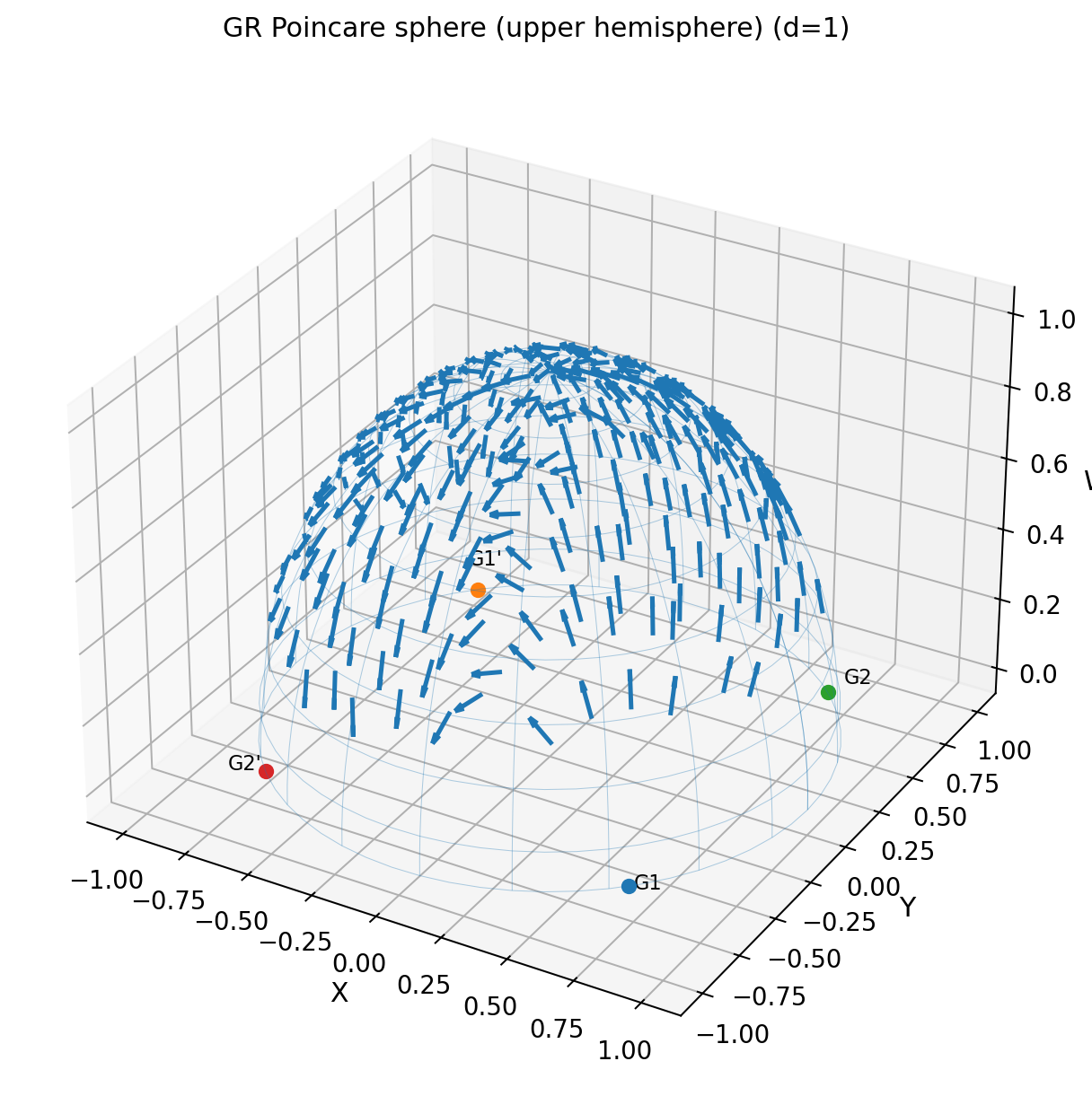}
    \includegraphics[width=0.400\textwidth]{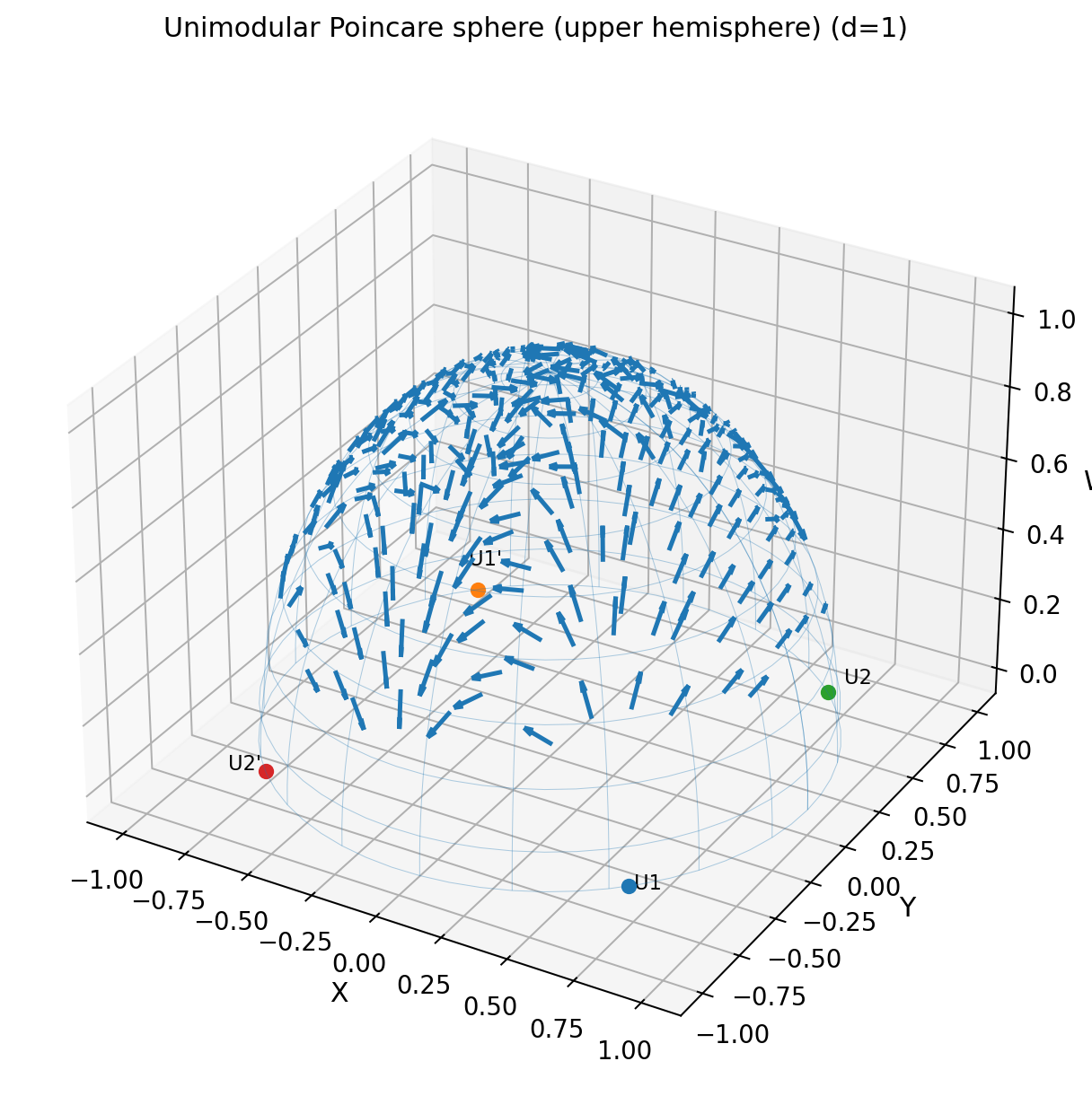}
    \caption{Poincar\'e disks and upper-hemisphere compactifications for the vacuum systems. The unimodular case exhibits a distinct asymptotic organization, showing that the extra-dimensional sector modifies not only the finite structure of the flow but also the global phase-space behavior at infinity.}
    \label{fig:vacuum_poincare}
\end{figure}

Figure~\ref{fig:vacuum_poincare} complements the finite phase portrait by showing the Poincar'e compactification of the vacuum flow. The disk and upper-hemisphere representations make explicit the asymptotic organization of the trajectories. In the unimodular case, the compactified flow is controlled by projective directions associated with the internal scalar sector. Notably, the analytical solutions presented in \cite{FabrisKerner2024,FabrisKerner2025} are represented by the vertical line crossing the unimodular Poincaré sphere, identifying them as persistent structures in the global phase space.

This confirms that the effect of the extra-dimensional unimodular contribution is not restricted to the neighborhood of finite critical points. The same geometric sector responsible for the equilibrium line in the finite plane also influences the preferred directions at infinity. The vacuum compactification therefore shows that the distinction with respect to the General Relativity is global in phase-space terms.

\subsection*{E. Coupled phase portrait and compactified flow}

Figures 5 and 6 summarize
the phase-space structure of the matter-coupled system in the
five-dimensional case $d=1$. Figure 5 displays the
finite reduced flow in the $(x,y)$ plane, where $x=\Omega_m$ and
$y=\lambda/H$. As expected from the analytical classification of
Sec.~IV.B, the trajectories are organized by the three critical points
$C_+$, $C_-$, and $M$, corresponding respectively to an attractor, a
repeller, and a saddle. The figure therefore gives a direct visualization
of how the reduced matter--geometry dynamics is structured at finite
distance.

The three-dimensional representation in the same figure should be read as
an auxiliary reconstruction of the Hubble evolution along the reduced
trajectories. The vertical coordinate is the normalized quantity
$E=H/H_0$, whose evolution is fixed once the trajectory in the $(x,y)$
plane is known. Thus this panel does not represent an independent
three-dimensional compactification; it only shows how the Hubble scale
changes along the finite reduced flow.

Figure 6 complements this picture by displaying
the Poincar\'e compactification of the same reduced system. The finite
critical points remain visible inside the disk, while the boundary
represents the asymptotic sectors of the flow. The preferred directions at
infinity are those obtained in Sec.~IV.B, namely the matter-free direction,
the direction with vanishing normalized internal-volume rate, and the
diagonal direction corresponding to an asymptotic balance between the
matter and geometric contributions.

\subsection*{F. Appendix A: Illustrative higher-dimensional phase portraits}

Figure 7 of the Appendix~A presents analogous phase-space portraits for the illustrative cases $d=2$ and $d=3$. These examples are included only to visualize how the reduced flow changes when the number of internal dimensions is varied. They are not intended as a full higher-dimensional critical-point analysis, which is left for future work. Rather, they provide a qualitative indication that the compactified phase-space organization found in the five-dimensional case persists in the representative examples considered.

\subsection*{G. General interpretation of the numerical results}

The numerical results presented in this section complement the analytical
phase-space analysis developed in Sec.~IV. Their purpose is not to add new
assumptions to the model, but to make explicit how the reduced equations
organize the cosmological evolution in representative situations. The
background comparison with $\Lambda$CDM shows that the model can reproduce
a nearby expansion history at low redshift while developing visible
departures as the internal-volume contribution evolves. This comparison is
therefore used only as a diagnostic benchmark, not as an observational
constraint.

The phase-space portraits provide the second part of this analysis. They
show, in visual form, the distinction between the vacuum reduced dynamics
and the minimally closed matter-coupled dynamics. In the vacuum case, the
figures emphasize the role of the internal-volume scalar in modifying the
global organization of the flow. In the matter-coupled case, the finite
and compactified portraits show how the normalized matter--geometry system
is structured both at finite distance and near the Poincar\'e boundary.
The examples in Appendix~A, for $d=2$ and $d=3$, are included only as
qualitative extensions of this picture, illustrating how the flow changes
when the number of internal dimensions is varied.

Therefore, the numerical section supports the main contribution of the
paper in three complementary ways: it connects the reduced equations with
background cosmological functions, displays the finite and asymptotic
phase-space organization of the model, and illustrates the persistence of
the compactified structure in representative higher-dimensional examples.
At the same time, the analysis remains deliberately limited in scope. A
full observational fit, a systematic treatment of arbitrary $d$, and the
study of perturbations are left for future work.

\begin{figure}[htbp]
    \centering
    \includegraphics[width=0.650\linewidth]{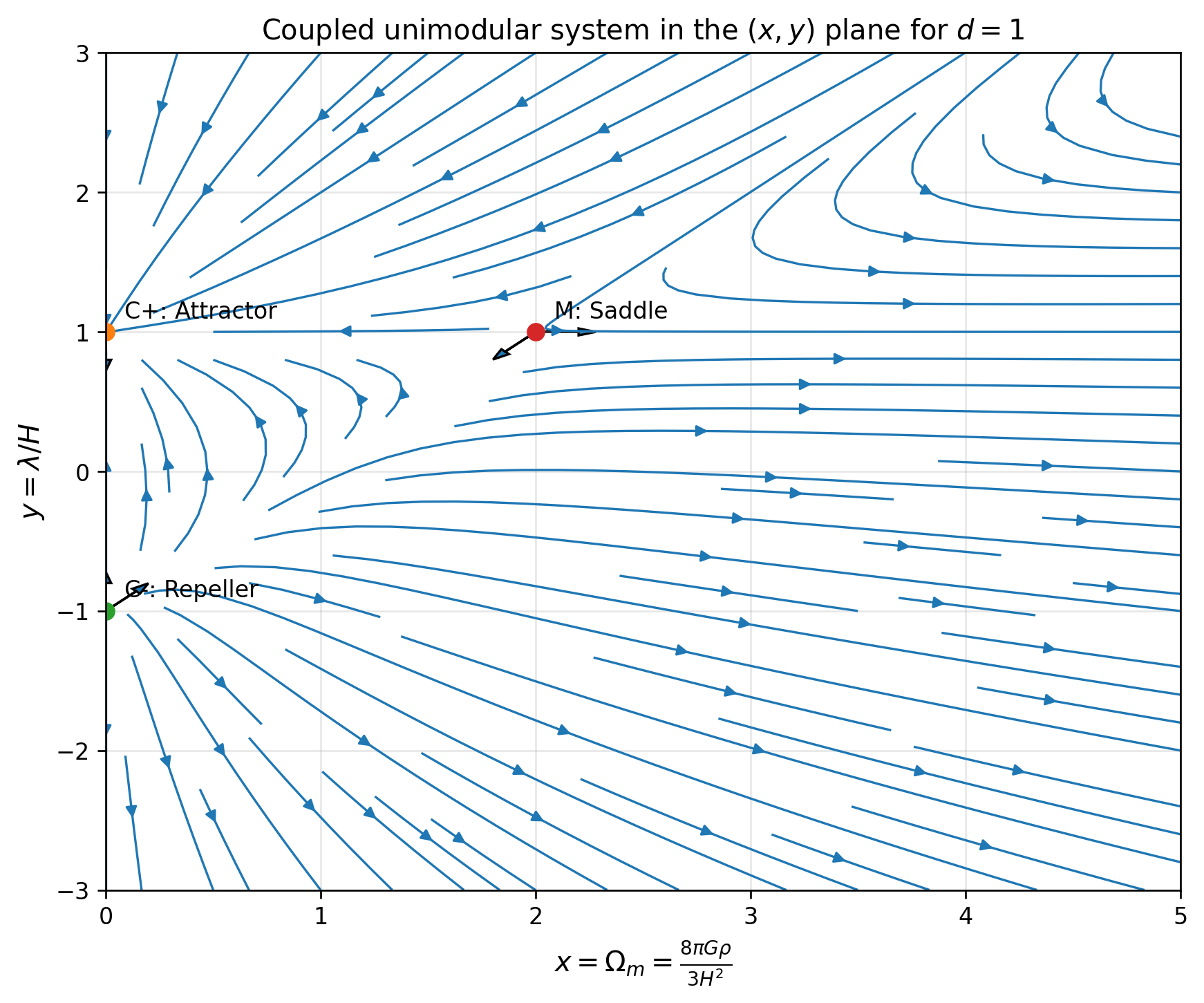}
    \includegraphics[width=0.650\textwidth]{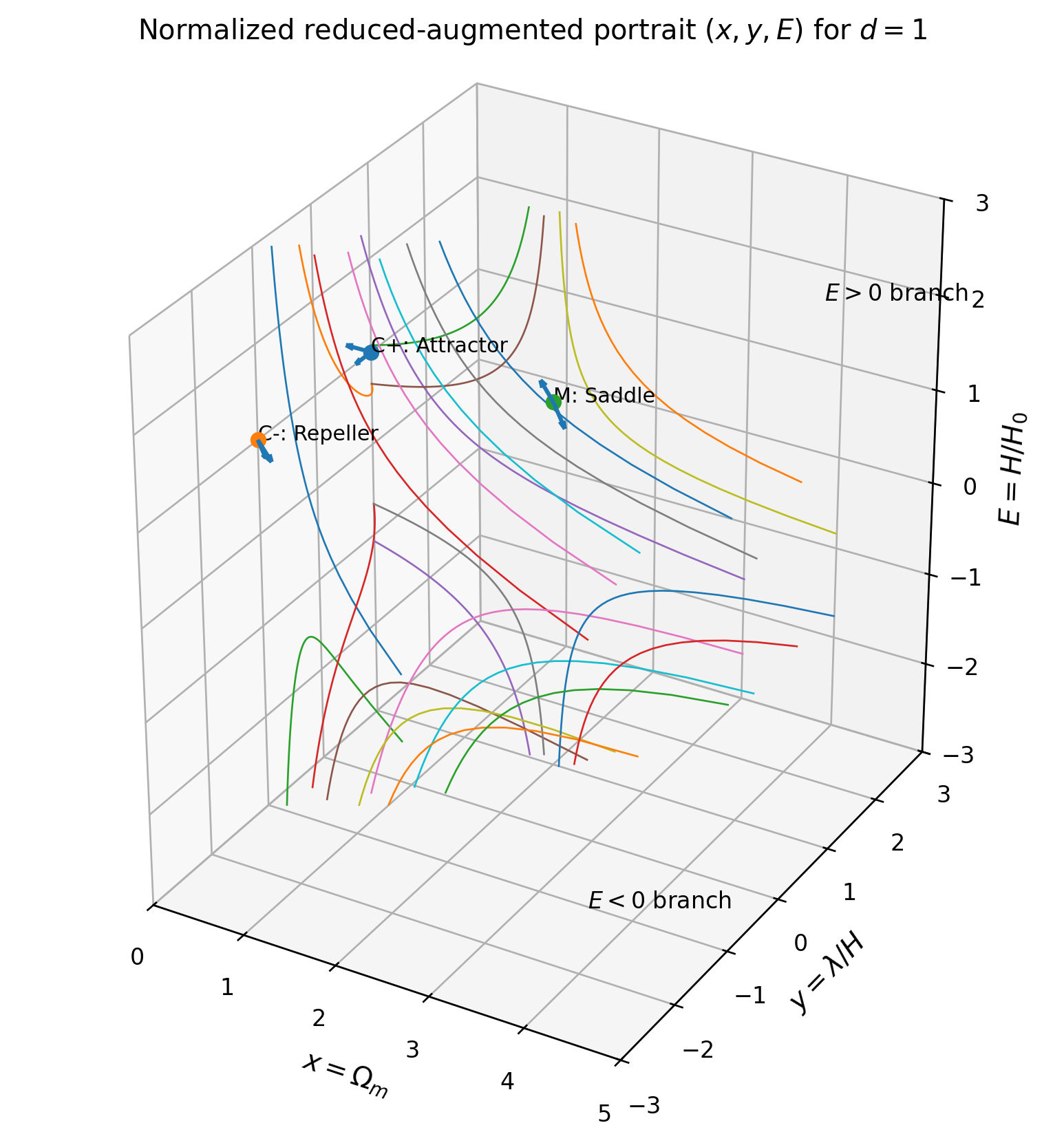}
    \caption{
Reduced phase portrait in the $(x,y)$ plane and $(x,y,H)$
representation for the coupled system with $d=1$, where
$x=\Omega_m=8\pi G\rho/(3H^2)$ and $y=\lambda/H$. The vector field is
computed from
$x'=2x^2-3xy-x$, $y'=2-x+xy-2y^2$, and $H'/H=y-x-1$.
The plotted ranges are $0\leq x\leq 5$, $-3\leq y\leq 3$, and
$-3\leq H\leq 3$, using a $45\times45$ grid in the reduced plane.
The  trajectories are integrated with step size $h=0.02$ over
$N_{\rm f}=N_{\rm b}=8$, with initial values
$H_0=\pm1.2$, $x_0=\{0.4,1.4,2.4,3.4\}$, and
$y_0=\{-1,0,1\}$. The reduced flow is organized by the three isolated
critical points $C_+=(0,1)$, $C_-=(0,-1)$, and the saddle
point $M=(2,1)$, corresponding respectively to the attractor, repeller,
and saddle sectors of the effective matter--geometry dynamics.
}
    \label{fig:coupled_phase_portrait}
\end{figure}

\begin{figure}[htbp]
    \centering
    \includegraphics[width=0.700\textwidth]{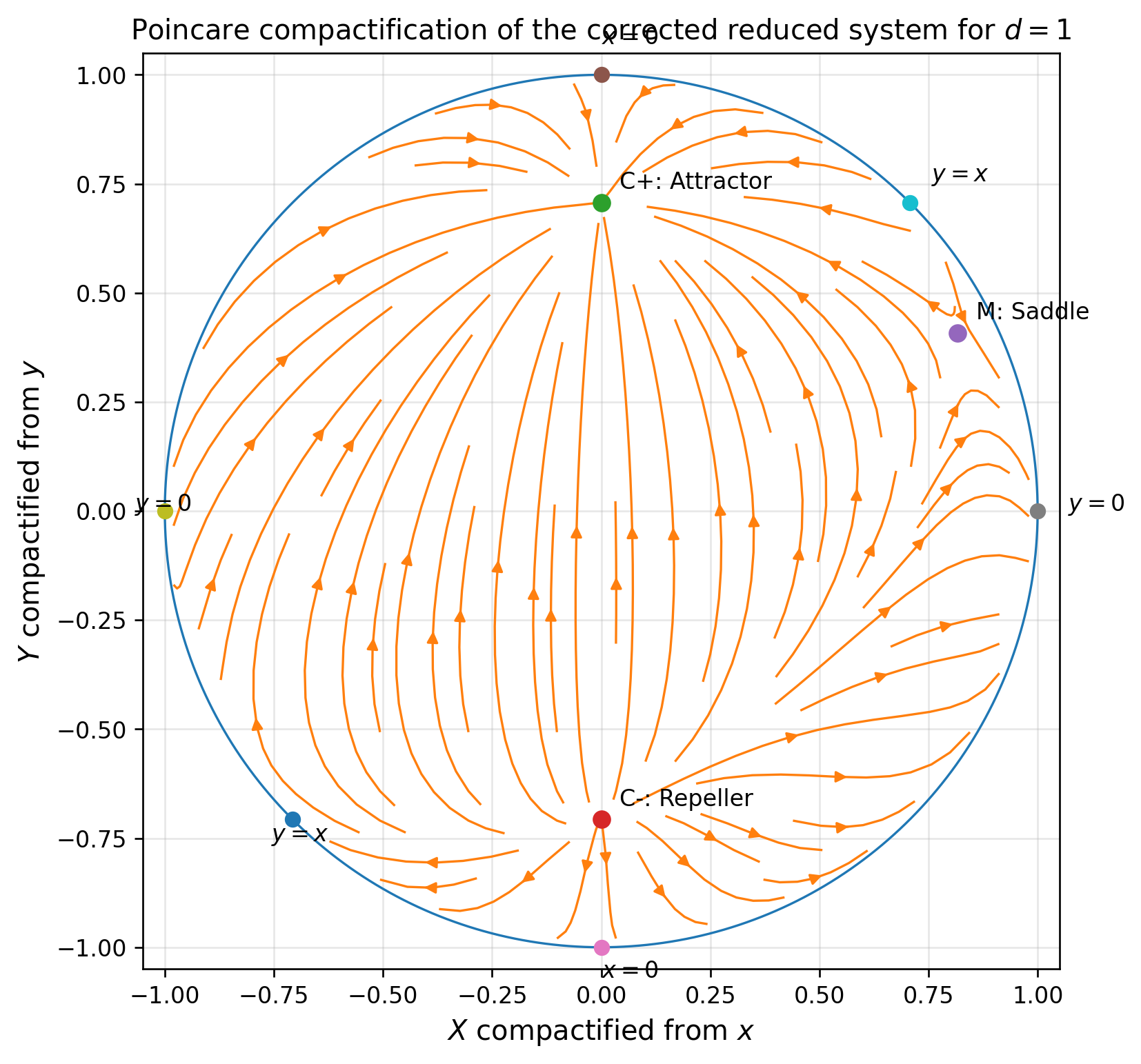}
   \caption{
Poincar\'e compactification of the reduced coupled system for
$d=1$, with $x=\Omega_m$ and $y=\lambda/H$. The compactified vector field
is obtained from the same reduced equations
$x'=2x^2-3xy-x$ and $y'=2-x+xy-2y^2$, using a $121\times121$ grid on
the unit disk. The finite critical points $C_+=(0,1)$, $C_-=(0,-1)$,
and $M=(2,1)$ are mapped into the compactified plane, while the boundary
directions are determined by the leading quadratic terms and correspond
to $x=0$, $y=0$, and $y=x$. The figure shows that the coupled unimodular dynamics remains
globally organized, with a well-defined attractor, repeller, saddle, and
asymptotic structure.
}
    \label{fig:coupled_poincare}
\end{figure}

\section{Conclusions}

We have investigated the effective four-dimensional cosmology obtained
from unimodular gravity in $D=4+d$ dimensions after dimensional reduction
of the internal-volume degree of freedom, $\psi=b^d$. The reduced FLRW
equations were formulated as autonomous systems, allowing us to analyze
both the finite critical structure and the global behavior of the
corresponding phase-space flows.

In the vacuum sector, the natural variables are the Hubble rate $H$ and
the internal-volume rate $\lambda=\dot{\psi}/\psi$. The main result is
that the reduced system does not contain only an isolated finite critical
configuration. Instead, besides the degenerate origin, it admits a
continuous equilibrium line, $\lambda=dH$. This line represents balanced
configurations in which the four-dimensional expansion and the evolution
of the internal volume remain dynamically correlated. The Poincar\'e
compactification further shows that the same ratio also appears as a
preferred asymptotic direction. Thus the internal-volume sector affects
both the local finite dynamics and the global organization of the vacuum
flow.

We then considered the matter-coupled sector in the five-dimensional case
$d=1$. The reduced consistency relation was used to identify the possible
matter--geometry exchange structure, and the system was closed by
adopting the minimal higher-dimensional conservation prescription. Under
this closure, matter is diluted by both the external spatial volume and
the internal-volume modulus. In the normalized variables
$x=\Omega_m$ and $y=\lambda/H$, the resulting reduced system has three
finite critical points: an attractor $C_+=(0,1)$, a repeller
$C_-=(0,-1)$, and a saddle point $M=(2,1)$. Therefore, in the minimally
closed matter sector, the continuous vacuum degeneracy is replaced by a
finite set of dynamically selected configurations.

The compactified matter flow has a simple asymptotic structure. The
directions at infinity are the matter-free boundary, the sector with
vanishing normalized internal-volume rate, and the diagonal direction
corresponding to an asymptotic balance between matter and geometry. This
shows that the reduced matter--geometry dynamics is organized not only by
its finite critical points, but also by a small number of preferred
asymptotic sectors on the Poincar\'e boundary.

The numerical examples were used as illustrations of the analytical
phase-space results. The comparison with $\Lambda$CDM shows that the
model can remain close to the standard background near the present epoch
while developing departures at higher redshift as the internal-volume
variable evolves. This comparison was intended only as a benchmark, not as
an observational fit. The additional portraits for $d=2$ and $d=3$ in
Appendix~A provide qualitative higher-dimensional visualizations of the
same construction, but they do not constitute a complete classification
for arbitrary $d$.

The analysis also makes clear the limitations of the present treatment.
The minimal closure used here fixes the four-dimensional Ricci scalar as
an integration constant and selects only one sector of the possible
matter--geometry exchange. More general exchange prescriptions would lead
to different reduced systems and must be studied separately. A complete
assessment of the model therefore requires extending the matter-coupled
critical-point analysis to arbitrary $d$, confronting the background
evolution with cosmological data, and studying linear perturbations in
order to determine whether the internal-volume degree of freedom leaves
observable signatures in structure formation.

\FloatBarrier
\clearpage
\appendix
\section{Numerical Results d=2 and d=3 }

In this appendix we present illustrative phase-space portraits for the
representative higher-dimensional cases $d=2$ and $d=3$. These examples
are not intended as a complete critical-point analysis for arbitrary
$d$. Their purpose is only to visualize how the reduced flow changes when
the number of internal dimensions is varied.

For this illustrative extension we use the pressureless matter equations
obtained from Eqs. (36) and (37), supplemented by the same minimal
higher-dimensional dilution law adopted in the main text,
\begin{eqnarray}
\dot{\rho}+3H\rho+\lambda\rho=0 .
\label{eq:appendix_minimal_closure}
\end{eqnarray}
Introducing the normalized variables
\begin{eqnarray}
x=\Omega_m=\frac{8\pi G\rho}{3H^2},
\qquad
y=\frac{\lambda}{H},
\qquad
N=\ln a ,
\label{eq:appendix_xy_def}
\end{eqnarray}
one obtains
\begin{eqnarray}
S_d(x,y)\equiv \frac{H'}{H}
=
-\frac{
3d^2+3dx+(1-d)y^2+d(d-4)y
}
{d(d+2)} .
\label{eq:appendix_Sd}
\end{eqnarray}
The reduced system used to generate the phase portraits is then
\begin{eqnarray}
x'
=
x\left[-3-y-2S_d(x,y)\right],
\label{eq:appendix_xprime_general}
\\
y'
=
\frac{
6d-3dx-3y^2+3(d-2)y
}
{d+2}
-yS_d(x,y).
\label{eq:appendix_yprime_general}
\end{eqnarray}
The normalized Hubble variable is reconstructed from
\begin{eqnarray}
E' = E S_d(x,y),
\qquad
E=\frac{H}{H_0}.
\label{eq:appendix_Eprime_general}
\end{eqnarray}

For $d=1$, Eqs.~\eqref{eq:appendix_Sd}--\eqref{eq:appendix_yprime_general}
reduce to the two-dimensional system analyzed in the main text,
\begin{eqnarray}
x'=2x^2-3xy-x,
\qquad
y'=2-x+xy-2y^2 .
\end{eqnarray}
For $d=2$ and $d=3$, the same formulas are used only to construct the
representative portraits displayed in Fig. 7.
Thus the appendix should be understood as a qualitative visualization of
higher-dimensional cases, rather than as a systematic classification of
the full matter-coupled dynamics for arbitrary $d$.

\begin{figure}[htbp]
    \centering
    \includegraphics[width=0.300\textwidth]{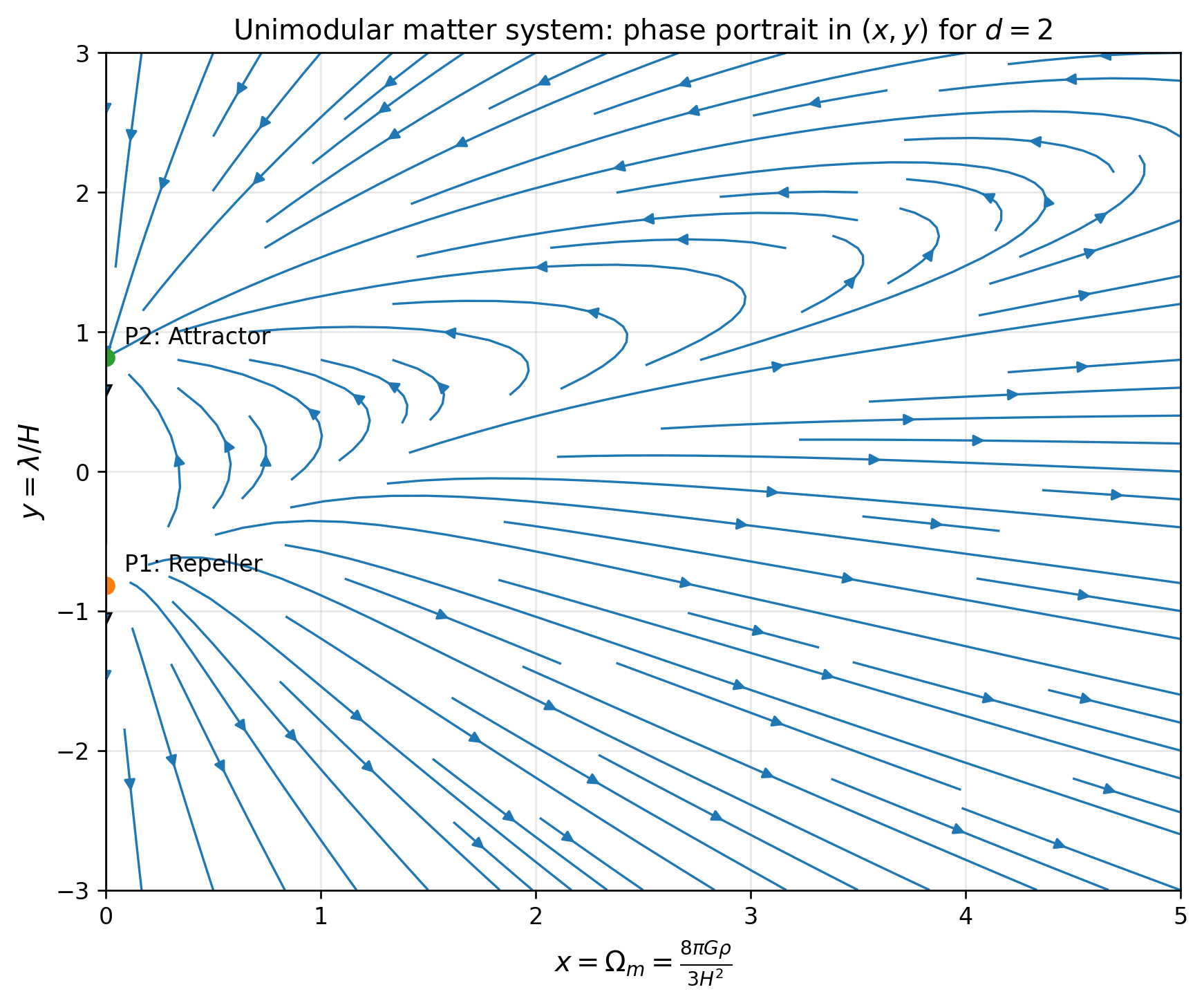}
    \includegraphics[width=0.300\textwidth]{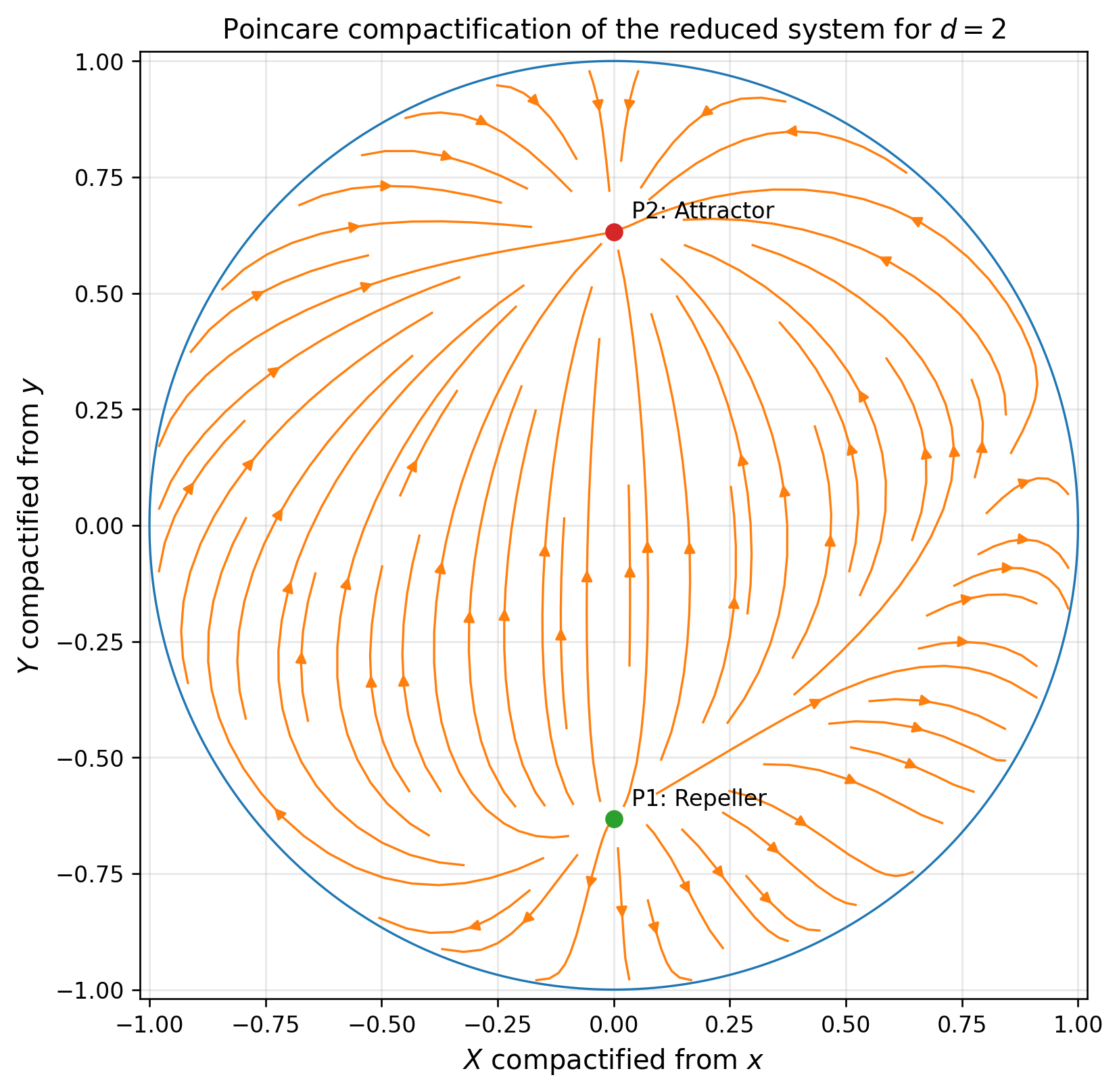}
    \includegraphics[width=0.300\textwidth]{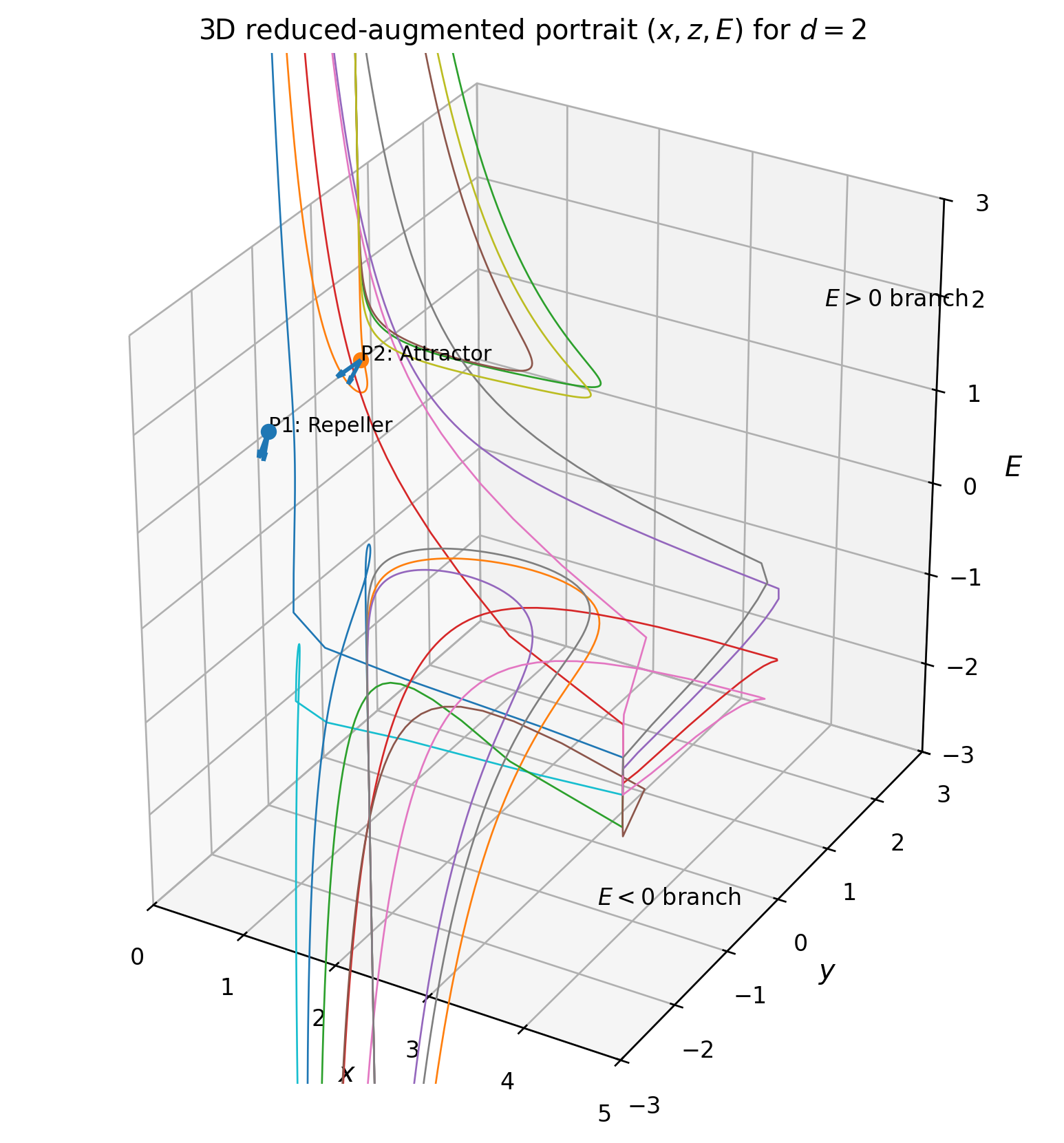}
    \includegraphics[width=0.300\textwidth]{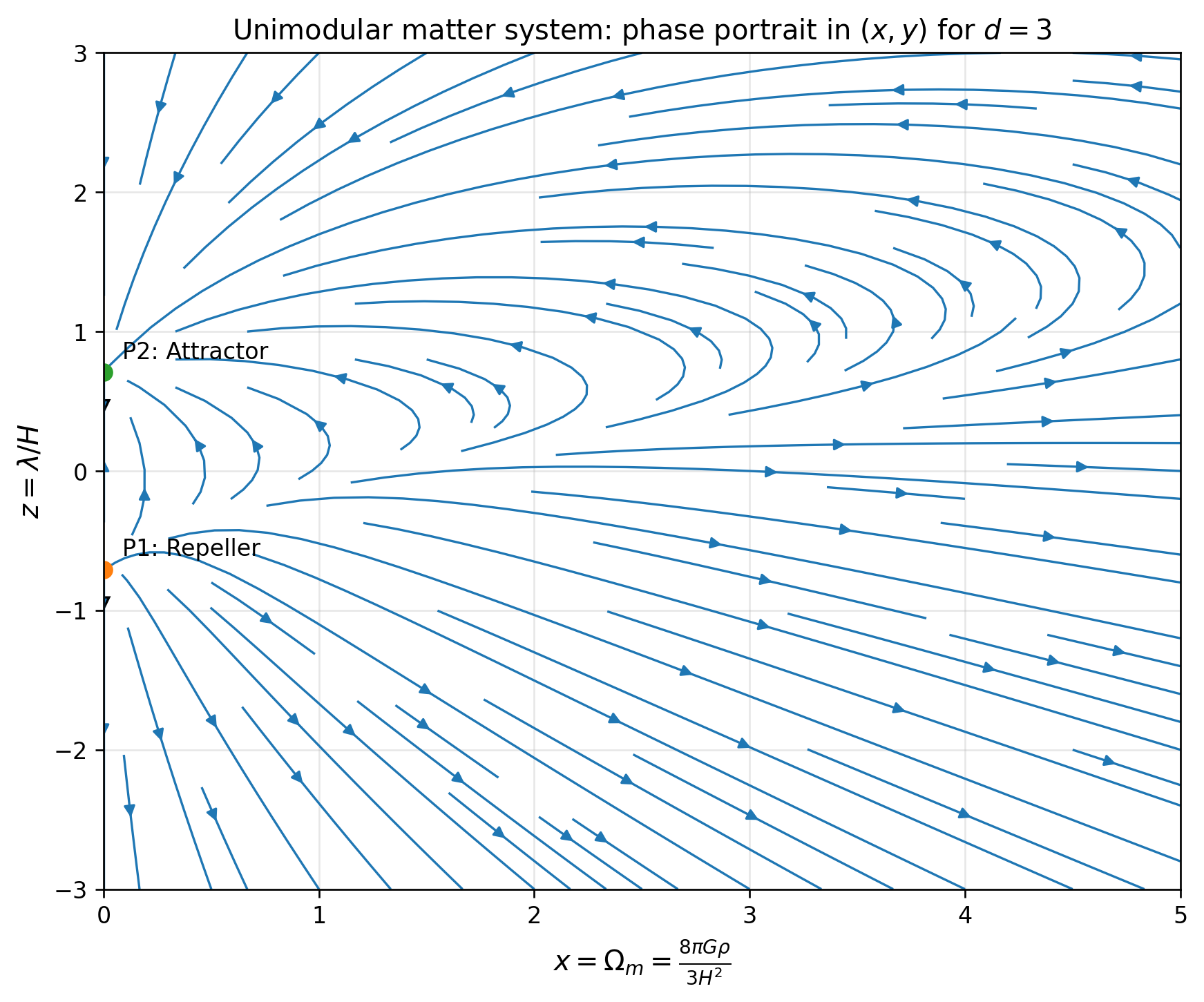}
    \includegraphics[width=0.300\textwidth]{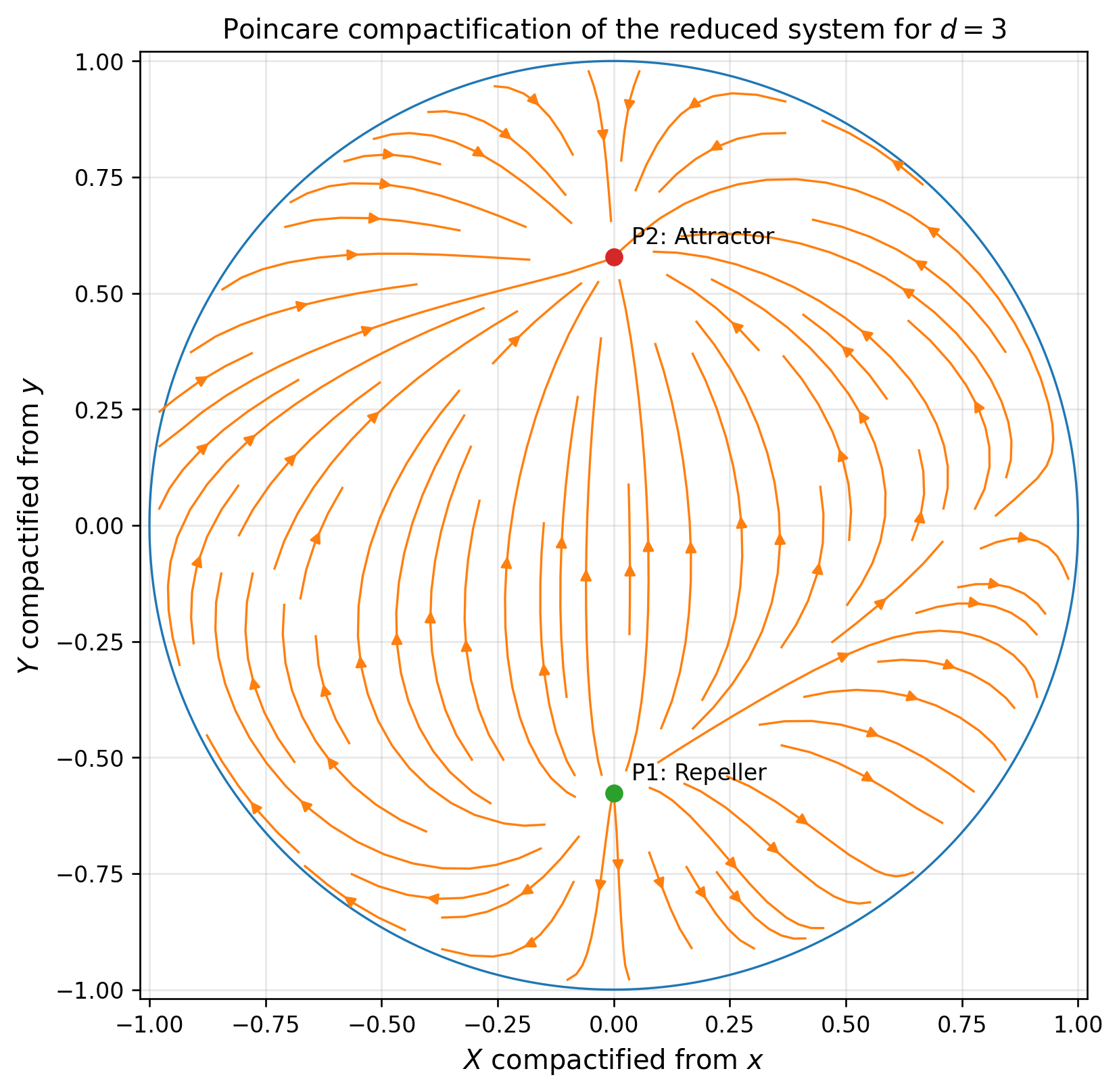}
    \includegraphics[width=0.300\textwidth]{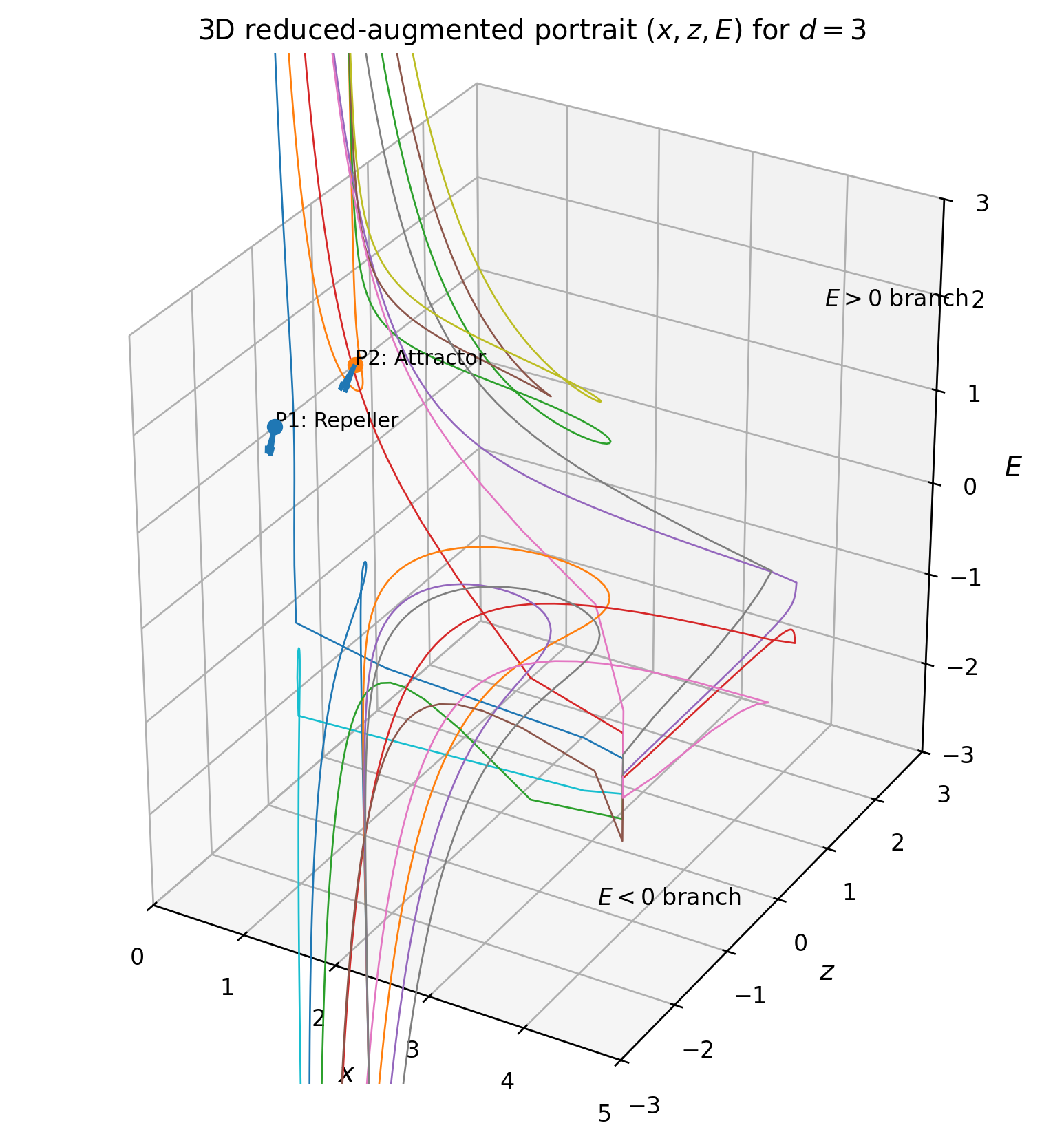}
    \caption{
Supplementary phase-space portraits for illustrative higher-dimensional
extensions with $d=2$ and $d=3$. For each value of $d$, the figure shows
the reduced phase portrait in the $(x,y)$ plane, the normalized
representation in $(x,y,E)$ with $E=H/H_0$, and the corresponding
Poincar\'e compactification. The variables are
$x=\Omega_m=8\pi G\rho/(3H^2)$ and $y=\lambda/H$, while $z$ is reserved
for the cosmological redshift. The numerical portraits use the ranges
$0\leq x\leq 5$, $-3\leq y\leq 3$, and $-3\leq E\leq 3$, with a
$45\times45$ grid in the reduced plane. The trajectories are
integrated with step size $h=0.02$ over
$N_{\rm f}=N_{\rm b}=8$, using initial values
$E_0=\pm1.2$, $x_0=\{0.4,1.4,2.4,3.4\}$, and
$y_0=\{-1,0,1\}$. The Poincar\'e disks are computed on a
$121\times121$ grid. The comparison shows that increasing the number of
internal dimensions produces only mild quantitative deformations of the
flow, while preserving the same qualitative phase-space organization in
the examples considered.
}
    \label{fig:coupled_poincare}
\end{figure}

\begin{acknowledgments}
A.M. Velásquez-Toribio thanks UFES for providing computational facilities to carry out this work, and J.C.F. thanks CNPq (Brazil) and FAPES (Brazil) for partial financial support. We also thank Richard Kerner for enlightening discussions during the development of this work.
\end{acknowledgments}

\end{document}